\documentclass[pra,twocolumn,showkeys,superscriptaddress,floatfix]{revtex4-1}

\pdfoutput=1

\usepackage[utf8]{inputenc}
\usepackage[english]{babel}
\usepackage{braket}
\usepackage{amsmath}
\usepackage{amsthm}
\usepackage{amssymb}
\usepackage{mathrsfs}
\usepackage{mathtools}
\usepackage{color}
\usepackage{comment}
\usepackage{multirow}
\usepackage{hepunits}
\usepackage[normalem]{ulem}
\makeatletter

\newcommand{\Rmnum}[1]{\expandafter\@slowromancap\romannumeral #1@}
\makeatother

\begin{document}

\title{Long-lifetime coherence in a quantum emitter induced by a metasurface}

\author{Emmanuel~Lassalle}
\email[]{emmanuel.lassalle@fresnel.fr}
\affiliation{Aix Marseille Univ, CNRS, Centrale Marseille, Institut Fresnel, F-13013 Marseille, France}

\author{Philippe~Lalanne}
\affiliation{LP2N, Institut d'Optique Graduate School, CNRS, Univ. Bordeaux, F-33400 Talence, France}

\author{Syed~Aljunid}
\affiliation{Centre for Disruptive Photonic Technologies, TPI \& SPMS Nanyang Technological University, 637371 Singapore}%

\author{Patrice~Genevet}
\affiliation{CRHEA, CNRS, Universit\'{e} C\^{o}te-d'azur, 06560 Valbonne, France}

\author{Brian~Stout}
%\email[]{brian.stout@fresnel.fr}
\affiliation{Aix Marseille Univ, CNRS, Centrale Marseille, Institut Fresnel, F-13013 Marseille, France}

\author{Thomas~Durt}
\affiliation{Aix Marseille Univ, CNRS, Centrale Marseille, Institut Fresnel, F-13013 Marseille, France}

\author{David~Wilkowski}
\email[]{david.wilkowski@ntu.edu.sg}
\affiliation{Centre for Disruptive Photonic Technologies, TPI \& SPMS Nanyang Technological University, 637371 Singapore}
\affiliation{Centre for Quantum Technologies, National University of Singapore, 117543 Singapore}
%\affiliation{MajuLab, CNRS-UCA-SU-NUS-NTU International Joint Research Unit, Singapore.}
\affiliation{MajuLab,~International~Joint~Research~Unit~UMI~3654,~CNRS,~Université Côte d’Azur,~Sorbonne Université, National University of Singapore, Nanyang Technological University, Singapore}

\date{\today}

\begin{abstract}
  An anisotropic quantum vacuum (AQV) has been predicted to induce quantum interferences during the spontaneous emission process in an atomic $V$-transition
  [G.~S.~Agarwal, Phys. Rev. Lett. \textbf{84}, 5500 (2000)].
  Nevertheless, the finite lifetime of the excited states is expected to strongly limit the observability of this phenomenon.
  In this paper, we predict that an AQV can induce a long-lifetime coherence in an atomic $\Lambda$-transition from the process of spontaneous emission, which has an additional advantage of removing the need for coherent laser excitation.
We also carry out two metasurface designs and compare their respective efficiencies for creating an AQV over remote distances.
The detection of this coherence induced by a metasurface, in addition to being yet another vindication of quantum electrodynamics,
could pave the way towards the remote distance control of coherent coupling between quantum emitters, which is a key requirement to produce entanglement in quantum technology applications.
\end{abstract}

\maketitle
%\tableofcontents

%\maketitle
%\tableofcontents
%\vskip20mm

\section{Introduction}

The control of the spontaneous emission of quantum emitters (QEs) has been investigated principally in a confined volume by the cavity-quantum electrodynamics (cQED) community \cite{raimond2006exploring}, whose archetype is a cavity formed by perfect mirrors.
The notion of ``cavity'' was then generalized to open resonators by the nanophotonics community \cite{baranov2017novel}, where strong couplings can be achieved. However, this typically  only occurs in the near-field of the photonic nanostructure and vanishes beyond a distance $d\sim \lambda_0$, where $\lambda_0$
is the emission wavelength of the QE in vacuum.
%This requires positionning the QE in very small volumes which can be quite challenging due to surface-interactions like Casimir-Polders interactions.
%Moreover, the lossy nature at optical wavelength in the case of metals make the quenching of the spontaneous emission very difficult to avoid...

There are a few other optical systems that can affect the spontaneous emission of QEs in the far-field ($d\gg\lambda_0$).
For instance, when covering half of the QE emission solid angle with a \emph{spherical mirror}, it has been predicted that the vacuum fluctuations can be 
fully suppressed at remote distances within a volume $\sim \lambda_0^3$, leading to a
total inhibition of the decay of a two-level atom \cite{hetet2010qed}.
In a classical picture, the field reflected by the spherical mirror can fully interfere with the direct field emitted by the atom:
if the atom is located at the focus of the spherical mirror such that $d=n\lambda_0/2$ with $n$ an integer number, there is a complete suppression
of the spontaneous emission, whereas if the atom is at the position $d=(n+1/2)\lambda_0/2$, the spontaneous emission is enhanced by a factor of 2.
Such effects occur provided that the round trip time of flight for the light to go from the
atom to the mirror and back is shorter  than the atom decay time $1/\gamma_0$ (with $\gamma_0$ the decay rate in free space),
that is for distances $d$ smaller than the \emph{photonic coherence length} $d_\text{cl}\equiv c/2\gamma_0$ \cite{dorner2002laser,kastel2005suppression}.
Such an alteration of the decay rate was already reported in Ref.~\cite{eschner2001light}, where the authors measured 1\%
change in the decay rate of an ion located at 30$\,$cm from a mirror. 

More recently, it has been suggested to use a reflecting metasurface acting as a spherical mirror to modify the spontaneous emission of a multilevel QE located at remote distances \cite{jha2015metasurface, jha2018spontaneous}.
This new paradigm unites the quantum optics and metasurface communities \cite{lalanne2017metalenses, genevet2017recent}, and relies on the fact that reflecting metasurfaces made of nano-resonators can break the isotropic nature of the vacuum to induce a polarization-dependent response, thus creating an anisotropic quantum vacuum (AQV).
It was previously predicted that an AQV can lead to quantum interferences in orthogonal levels of a multilevel QE in a $V$-configuration, that is two excited states and one ground state \cite{agarwal2000anisotropic}. 
However, the predicted effects, \emph{i.e.} a population transfer between the two excited states of $\sim 1$\% \cite{jha2015metasurface}, and
an induced coherence of about 10\% \cite{jha2018spontaneous}, only last as long as the atom remains in its excited states,
which is a drastic drawback for experimental confirmations.

Although the $V$-scheme is the one most often considered in the literature  \cite{agarwal2000anisotropic,agarwal2001vacuum,li2001quantum,yang2008quantum,jha2015metasurface,sun2016quantum,hughes2017anisotropy,jha2018spontaneous},
this work focuses on the spontaneous emission properties of a QE with a $\Lambda$-transition, i.e. a single excited state linked to two nearly degenerate ground states,
in an AQV created by a metasurface. We predict the generation of a coherence between
the two ground states, which survives after the photon emission. %, in contrast with the $V$-scheme. % proposed in earlier works \citep{agarwal2000anisotropic,jha2015metasurface,jha2018spontaneous}.
The interest in the ground state coherence arises from its long lifetime, which allows high resolution experiments.
Moreover, it was previously known that this coherence could only be generated with an external coherent laser field (see \cite{suter1997physics}, Chapter 3).
Here, we show that such a coherence can be simply generated by spontaneous emission in an anisotropic vacuum (in the absence of a laser field).

In Section~\ref{ch4:section:theory}, we derive the master equation for the $\Lambda$-scheme (Section~\ref{sec:me}),
and we show how an anisotropic vacuum can induce a coherence between the ground states
from the process of spontaneous emission (Section~\ref{sec:aqvvvv}).
We also provide an interpretation of this result in terms of the dressed-states of the system (Section~\ref{sec:dressed-states}).
In Section~\ref{ch4:section:meta}, following a phase-mapping approach (presented in Section~\ref{sect:pma}), we propose two designs of metasurfaces to realize the anisotropic vacuum and characterize their performances (Sections~\ref{sec:resonant-phase-delay} and \ref{sec:geometa}).
Finally, we assess the value of the coherence that can be achieved using such metasurfaces, taking into account the limitations due to the finite size of their nano-resonators (Section~\ref{ch4:sec:limitations}).

\section{Theoretical prediction: long lifetime coherence}
\label{ch4:section:theory}

\begin{figure}[b]
   \centering
   \includegraphics[scale=0.5]{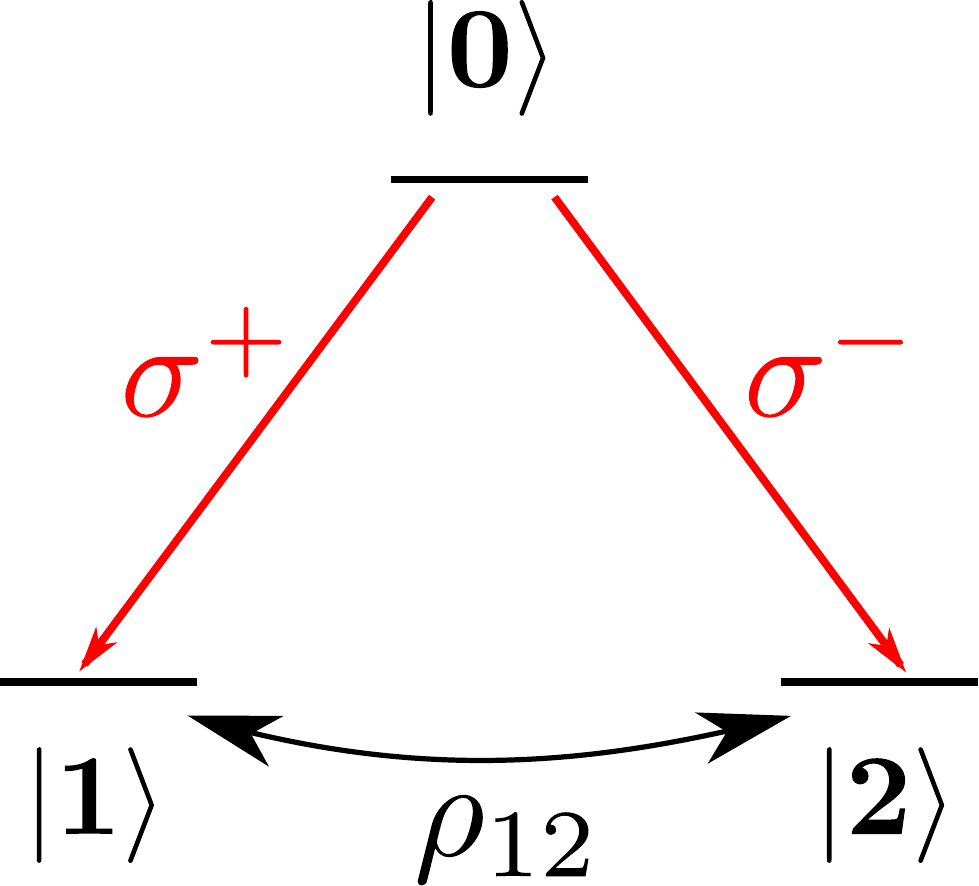}
   \caption{Three-level quantum emitter with a $\Lambda$-structure. The upper level $\ket{0}$ can decay 
     \emph{via} two transitions: either to the state $\ket{1}$ with the emission of a right circularly polarized photon denoted $\sigma^+$, or to the state $\ket{2}$ with the emission of a left circularly
     polarized photon denoted $\sigma^-$. $\rho_{12}$ denotes the coherence between the two ground states $\ket{1}$ and $\ket{2}$.}
   \label{fig:structure}
 \end{figure}

We consider a three-level system in a so-called $\Lambda$-scheme: one single excited state $\ket{0}$, which can decay into two ground states $\ket{1}$ and $\ket{2}$ \emph{via} two
orthogonal dipolar transitions by the emission of circularly polarized photons $\sigma^+$ and, respectively, $\sigma^-$ (see Fig.~\ref{fig:structure}).
By orthogonal transitions, it means that the dipole moments $\bold{d}_{01}$ and $\bold{d}_{02}$ corresponding to these transitions are orthogonal (\emph{i.e.} $\bold{d}_{01}^*\cdot \bold{d}_{02}=0$).
They are given by: 
$\bold{d}_{01}=+d_{01}\vec{\varepsilon}_+$ and $\bold{d}_{02}=-d_{02}\vec{\varepsilon}_-$ where $\vec{\varepsilon}_\pm=(\vec{x}\pm\mathrm{i}\vec{y})/\sqrt{2}$. We use the $\vec{z}$ direction as the quantization axis.
This scheme appears naturally in NV-centers in diamond, using the magnetic sublevels $|\pm1\rangle$ as the ground states and $|A_2\rangle$ as the excited state \cite{togan2010quantum}.
It also can be found in atoms, using Zeeman manifold with $|F,m=\pm1\rangle$ for the ground states and $|F',m=0\rangle$ for the excited state, where
$m$ are the magnetic quantum numbers, and $F$, $F'$ are the total angular momentum quantum numbers \cite{suter1997physics}.

The interaction between the atom (at position $\bold{r}_0$) and the electromagnetic (EM) environment in the vacuum state (\emph{i.e.} no photons) is described by the
interaction Hamiltonian in the \emph{electric dipole approximation}:
$\hat{H}_I = -\hat{\mathbf{d}}\cdot\hat{\mathbf{E}}_v(\bold{r}_0)$.
The dipole moment operator $\hat{\mathbf{d}}$ is given by: $\hat{\mathbf{d}} = \mathbf{d}_{01}\ket{0}\bra{1}+\mathbf{d}_{02}\ket{0}\bra{2} + \mathbf{d}_{01}^*\ket{1}\bra{0}+\mathbf{d}_{02}^*\ket{2}\bra{0}$.
The electromagnetic field operator
$\hat{\mathbf{E}}_v$ (where the subscript $v$ denotes the vacuum) can formally be written as a sum of a complex field
$\hat{\mathbf{E}}_v^{(+)}$ and its Hermitian conjugate (H.c.)
$\hat{\mathbf{E}}_v^{(-)}=[\hat{\mathbf{E}}_v^{(+)}]^\dagger$: $\hat{\mathbf{E}}_v(\bold{r}_0)=\hat{\mathbf{E}}_v^{(+)}(\bold{r}_0)+\hat{\mathbf{E}}_v^{(-)}(\bold{r}_0)$.
In the interaction picture (time-dependent Hamiltonian), and after making the \emph{rotating wave approximation}, this interaction Hamiltonian reads:
%\begin{equation}
  \begin{multline}
\hat{H}_I(t) = \\-\left(\mathbf{d}_{01}\ket{0}\bra{1}e^{\mathrm{i}\omega_1 t}+\mathbf{d}_{02}\ket{0}\bra{2}e^{\mathrm{i}\omega_2 t}\right)\cdot\hat{\mathbf{E}}_v^{(+)}(\mathbf{r}_0,t)\\
-\left(\mathbf{d}_{01}^*\ket{1}\bra{0}e^{-\mathrm{i}\omega_1 t}+\mathbf{d}_{02}^*\ket{2}\bra{0}e^{-\mathrm{i}\omega_2 t}\right)\cdot\hat{\mathbf{E}}_v^{(-)}(\mathbf{r}_0,t)\; ,
%\end{split}
\label{chaqv:intH}
\end{multline}
where $\omega_i$ is the transition frequency associated
with the transition $\ket{0}\rightarrow \ket{i}$
($i=1,2$).
We derive in Section~\ref{sec:me} the master equation for the reduced density matrix of the atom.

%\begin{comment}
 %\end{comment}

\subsection{Master equation for the atomic density matrix}
\label{sec:me}

The total system \{atom + EM environment\} is characterized by the density matrix $\rho_T(t)$,
which obeys the Schr\"odinger equation which, in the interaction picture, reads \cite{barnett2002methods,carmichael2013statistical}:
\begin{equation}
\frac{\partial \rho_T(t)}{\partial t} =
\frac{1}{\mathrm{i}\hbar}[\hat{H}_I(t),\rho_T(t)]\; ,
\label{eq:rho_tot}
\end{equation}
with $\hat{H}_I(t)$ given by Eq.~(\ref{chaqv:intH}).
The reduced density matrix of the atom (atomic density matrix) is obtained by taking the trace over
the degrees of freedom of the environment: $\rho(t) \equiv \text{Tr}_e(\rho_T(t))$. 
In order to find the master equation governing the evolution of this reduced density matrix, we first assume that there is no correlation between the atom and the EM environment
at time $t=0$, so that $\rho_T(0)$ factorizes as:
$\rho_T(0)=\rho(0)\otimes\rho_e(0)$, with $\rho_e$ the reduced density matrix of the EM environment. Moreover, considering that only the state of the atom is affected by the interaction,
we assume that at later times $t$, $\rho_T(t)$ factorizes as: $\rho_T(t)=\rho(t)\otimes\rho_e(0)$.
Finally, by making two other major approximations, known as the \emph{Born and Markov approximations},
we obtain the following master equation for the atomic density matrix $\rho(t)$ (where we have considered for simplicity closed-lying states: $\omega_1\simeq\omega_2\equiv\omega_0$;
see Appendix for the details of the derivation):
\begin{widetext}
\begin{equation}
  \frac{\partial \rho(t)}{\partial t} = -\left[\mathrm{i}\omega_0+\frac{\gamma_1}{2}+\frac{\gamma_2}{2}\right]\ket{0}\bra{0}\rho(t)
  +\rho_{00}(t)\left[\frac{\gamma_1}{2}\ket{1}\bra{1}+\frac{\gamma_2}{2}\ket{2}\bra{2}+\frac{\kappa_{21}}{2}\ket{2}\bra{1}
+\frac{\kappa_{12}}{2}\ket{1}\bra{2}\right]
+\text{H.c.}
\label{chaqv:eq:master_eq_main}
\end{equation}
\end{widetext}
In Eq.~(\ref{chaqv:eq:master_eq_main}), $\rho_{00}(t)$ denotes the population in the excited state $\ket{0}$ (defined as $\rho_{00}(t)\equiv \bra{0}\rho(t)\ket{0}$),
and we have introduced the coefficients $\gamma_i$ and $\kappa_{ij}$, whose expressions are:
\begin{equation}
\gamma_i \equiv \frac{1}{\hbar^2}\bold{d}_{0i}^*\cdot
\hat{\mathbf{C}}(\mathbf{r}_0,\mathbf{r}_0,\omega_0)\cdot\bold{d}_{0i} \quad (i=1,2) \; ,
\label{eq:gamma_first}
\end{equation}
and 
\begin{equation}
\kappa_{12} \equiv \frac{1}{\hbar^2}\bold{d}_{01}^*\cdot
\hat{\mathbf{C}}(\mathbf{r}_0,\mathbf{r}_0,\omega_0)\cdot\bold{d}_{02} \; .
\label{eq:kappa_main_text}
\end{equation}
The coefficient $\gamma_i$ characterizes the transition $\ket{0}$ to $\ket{i}$, and is called \emph{decay rate}; the coefficient
$\kappa_{12}$ characterizes a cross-coupling between the states $\ket{1}$ and $\ket{2}$ (and $\kappa_{21}=\kappa_{12}^*$).
These coefficients are defined in Eqs.~(\ref{eq:gamma_first}) and (\ref{eq:kappa_main_text}) in terms of the \emph{correlation tensor} $\hat{\mathbf{C}}$:
\begin{equation}
\hat{\mathbf{C}}(\mathbf{r},\mathbf{r}',\omega)\equiv\int_{-\infty}^{+\infty}\mathrm{d}\tau\,\left<\hat{\mathbf{E}}_v^{(+)}(\mathbf{r},\tau)\hat{\mathbf{E}}_v^{(-)}(\mathbf{r}',0)\right>e^{\mathrm{i}\omega\tau}\; ,
\label{chaqv:eq:correltenspossss}
\end{equation}
where the bracket indicates an ensemble average:\\
%\begin{equation}
$\left<\hat{\mathbf{E}}_v^{(+)}(\mathbf{r},\tau)\hat{\mathbf{E}}_v^{(-)}(\mathbf{r}',0)\right>~\equiv~\text{Tr}_e\left(\rho_e(0)\hat{\mathbf{E}}_v^{(+)}(\mathbf{r},\tau)\hat{\mathbf{E}}_v^{(-)}(\mathbf{r}',0)\right)$ (see Appendix).
%\end{equation}
This correlation tensor characterizes the amplitude of the \emph{fluctuations} of the electric field in the vacuum state, which contain all the information about the dynamics of the system since,
once they are known, the dynamics of the atom given by Eq.~(\ref{chaqv:eq:master_eq_main}) can in principle be solved.

We now integrate Eq.~(\ref{chaqv:eq:master_eq_main}) for an atom initially prepared in the excited state corresponding to the following initial conditions at $t=0$:
$\rho_{00}(0)=1$, $\rho_{11}(0)=\rho_{22}(0)= 0$ and $\rho_{ij}(0)=0$ for $j\neq i$, where $\rho_{ii}(t)$ is the atomic population in the state $\ket{i}$ and $\rho_{ij}(t)$ is the atomic coherence between the states $\ket{i}$ and $\ket{j}$.
For the steady state ($t\rightarrow \infty$), we find, for the atomic populations, that $\rho_{00}(\infty) = 0$ and
%\begin{equation}
%\rho_{00}(\infty) = 0
%\label{eq:rho_33_bis}
%\end{equation}
\begin{equation}
\rho_{ii}(\infty) = \frac{\gamma_i}{\gamma_1+\gamma_2}\quad (i=1,2)\; ,
\label{ch4:eq:pop1}
\end{equation}
and, for the atomic coherences, that $\rho_{10}(t)=\rho_{20}(t)=0$ ($\forall t$) and (using the fact that $\kappa_{21}^*=\kappa_{12}$)
%\begin{equation}
%\rho_{10}(t)=\rho_{20}(t)=0 \quad \forall t
%\end{equation}
\begin{equation}
\rho_{12}(\infty)=\frac{\kappa_{12}}{\gamma_1+\gamma_2}\; .
  \label{ch4:eq:coh_infty}
\end{equation}

While the result in Eq.~(\ref{ch4:eq:pop1}) simply shows that the populations in the steady state are in a probabilistic distribution either in state $\ket{1}$ or $\ket{2}$,
the result in Eq.~(\ref{ch4:eq:coh_infty}) for the coherence $\rho_{12}$ is more surprising:
it reveals that a coherence between the two ground states can be induced by
\emph{spontaneous emission}, \emph{i.e.  without} an external field; while to date, it was thought that a coherence between the two ground states required an external coherent field such as a laser field (Ref.~\cite{suter1997physics}, Chapter 3).
Furthermore, because it involves ground states, this coherence has in principle a long lifetime, in the millisecond range for NV-centers at room temperature \cite{balasubramanian_ultralong_2009},
  and in the order of seconds for cold atom systems where collisions are suppressed. Therefore, we have simply ignored the relaxation term of $\rho_{12}$ in Eq.~(\ref{chaqv:eq:master_eq_main}),
  which is supposed to be of much longer time than the coherence involving the excited state.

  The detection of the coherence $\rho_{12}$ in NV-centers can be performed following the protocols discussed in Ref.~\cite{togan2010quantum}: a magnetic field bias
  is applied to lift the degeneracy between the two ground states $|\pm1\rangle$. It allows one to address separately the transitions $|\pm1\rangle\rightarrow|0\rangle$ with two
  microwave fields, where $\ket{0}$ is another magnetic sublevel. The presence of coherence between the state $|\pm1\rangle$ results in a phase-sensitive transfer to $|0\rangle$.
  Finally, the population of $|0\rangle$ is probed optically using a cycling transition with an auxiliary excited state $\ket{E_y}$ (see Ref.~\cite{togan2010quantum}).
  With cold atomic ensembles, a similar method could be employed using the hyperfine structure of the ground state of Alkali-metal atoms.
  In this case, the coherence $\rho_{12}$ is generated between two states within one Zeeman manifold.
  Then, two radiofrequency fields perform a phase-sensitive transfer to a state belonging to another Zeeman manifold. Finally, the population of this last state is optically measured.
  %An alternative way could be to implement a tomography technique \cite{PhysRevA.64.052312}.

\subsection{Anisotropic quantum vacuum}
\label{sec:aqvvvv}

We will now find the conditions for the existence of the long-lifetime coherence of Eq.~(\ref{ch4:eq:coh_infty}).
For that, we first use the \emph{fluctuation-dissipation theorem} at zero temperature (we do not consider the effect of the temperature, which is indeed very small when one considers
an atom emitting at optical frequencies). This theorem links the correlation tensor $\hat{\mathbf{C}}$ of Eq.~(\ref{chaqv:eq:correltenspossss}), which we recall characterizes the vacuum electric field \emph{fluctuations},
to the imaginary part of the Green tensor $\hat{\mathbf{G}}$, which describes the \emph{dissipation} of the electric energy, as \cite{agarwal2000anisotropic}:
\begin{equation}
\hat{\mathbf{C}}(\bold{r},\bold{r}',\omega) =
\frac{2\hbar\omega^2}{\epsilon_0c^2}\text{Im}\left(\hat{\mathbf{G}}(\bold{r},\bold{r}',\omega)\right)\; .
\end{equation}
The fluctuation-dissipation theorem shows that the amplitude of the fluctuations are known once the imaginary part of the Green tensor has been calculated.
Making use of it, the coefficients $\gamma_i$ [Eq.~(\ref{eq:gamma_first})] and $\kappa_{12}$ [Eq.~(\ref{eq:kappa_main_text})] can be expressed in term of the Green tensor as:
\begin{equation}
\gamma_i = \frac{2\omega_0^2}{\hbar\epsilon_0 c^2}\,\bold{d}_{0i}^*\cdot
\text{Im}\left(\hat{\mathbf{G}}(\bold{r}_0,\bold{r}_0,\omega_0)\right)\cdot\bold{d}_{0i}\; ,
\label{ch4:eq:gamma_Green}
\end{equation}
and
\begin{equation}
\kappa_{12} = \frac{2\omega_0^2}{\hbar\epsilon_0 c^2}\,\bold{d}_{01}^*\cdot
\text{Im}\left(\hat{\mathbf{G}}(\bold{r}_0,\bold{r}_0,\omega_0)\right)\cdot\bold{d}_{02} \; .
\label{ch4:eq:kappa}
\end{equation}
Next, we express the Green tensor and the dipole moments appearing in Eqs.~(\ref{ch4:eq:gamma_Green}) and (\ref{ch4:eq:kappa})
in the Cartesian basis $(\vec{x},\vec{y},\vec{z})$ (we recall that a static magnetic field is applied along the $\vec{z}$ direction, defining the quantization axis). Eq.~(\ref{ch4:eq:coh_infty}) can then be recast in the following form (using the fact that $G_{yx}=G_{xy}$):
\begin{equation}
\rho_{12}(\infty)=\underbrace{\frac{d_{01}d_{02}}{d_{01}^2+d_{02}^2}}_R\times\underbrace{\frac{\text{Im}\left[G_{xx}-G_{yy}\right]-\mathrm{i}2\text{Im}\left[G_{xy}\right]}{\text{Im}\left[G_{xx}+G_{yy}\right]}}_A\; ,
\label{ch4:rhocartesian}
\end{equation}
where the Green tensor Cartesian components have to be evaluated at the position of the quantum emitter $\bold{r}_0$ and at the transition frequency $\omega_0$.

One immediately remarks that in the usual isotropic vacuum, $G_{xx}=G_{yy}$ and $G_{xy}=0$, and Eq.~(\ref{ch4:rhocartesian}) predicts null coherence.
Therefore, in order to generate coherence, %when the dipole matrix elements are orthogonal,
the vacuum has to be \emph{anisotropic}. A similar result was first put forward by G.S.~Agarwal in
Ref.~\cite{agarwal2000anisotropic} for a $V$-configuration,
where he predicted a coherent population transfer between the two orthogonal excited states in an AQV. 

To quantify the anisotropy, the coherence in Eq.~(\ref{ch4:rhocartesian}) can be written as a product of two terms: the coefficients $R$ and $A$, characterizing the quantum emitter on one hand,
and the vacuum anisotropy on the other hand. $R$ reaches its maximum value of $0.5$ when the two dipole moment amplitudes are equal ($d_{01}=d_{02}$).
The coefficient $A$ (refered to hereafter as the ``anisotropy''), in its general form, is a complex quantity, and depends on the EM environment which is completely characterized by $\hat{\mathbf{G}}$.
In this work, we will only consider situations where $G_{xy}=0$ (which will be justified later),
%Since the two metasurface designs, discussed in Section \ref{ch4:section:meta}
%are made of nano-resonators that respect the mirror symmetry, $G_{xy}=0$.
so from now on, $A$ will be considered
as a real quantity and takes the form of a \emph{visibility} with extremum values $\pm 1$.
Therefore the extrema of the coherence are $\rho_{12}(\infty)=\pm 1/2$.
%In Fig.~\ref{ch4:multilevelatom}, we plot this coherence as a function of time [derived in Appendix~\ref{chaqv:app:me}, Eq.~(\ref{app:ch4:eq:coh_time})] in the case where $\kappa_{12}=\gamma_1=\gamma_2=\gamma_0/2$
%where $\gamma_0$ is the decay rate of a \emph{two-level atom} of dipole moment amplitude $d$ in free space (considering same dipole moment amplitudes $d=d_{01}=d_{02}$). %The upper bound of $0.5$ is derived in the next Section.
\begin{comment}
\begin{figure}[h!]
   \centering
   \includegraphics[width=\linewidth]{./coherence.pdf}
   \caption{Coherence $\rho_{12}(t)$ [Eq.~(\ref{app:ch4:eq:coh_time}) in Appendix~\ref{chaqv:app:me}] between the ground states $\ket{1}$ and $\ket{2}$ (see Fig.~\ref{fig:structure}) in an anisotropic quantum vacuum, as a function of the (normalized) time $\gamma_0 t$,
     in the case $\kappa_{12}=\gamma_1=\gamma_2=\gamma_0/2$, where $\gamma_0$ is the decay rate
     of a two-level atom of dipole moment $d$ in free space (considering same dipole moments $d=d_{01}=d_{02}$).}
   \label{ch4:multilevelatom}
\end{figure}
\end{comment}

\subsection{Interpretation in terms of dressed-states}
\label{sec:dressed-states}

In situations where the coherence is extremum, the atomic density matrix, after spontaneous emission, reads in the basis of the two ground states \{$\ket{1},\ket{2}$\}:
$\rho(\infty)=\frac{1}{2}
\begin{bmatrix}
1 & \pm1\\
\pm1 & 1 \\
\end{bmatrix}$,
which corresponds to a pure state. This is in stark contrast with the \emph{isotropic} vacuum where spontaneous emission produces  a statistical mixture with a reduced density matrix $\rho(\infty)=\frac{1}{2} \mathbb{I}$.
%One can find a discussion in terms of the dressed-state of the system in Appendix~\ref{app:dressed-state}.

One can interpret this in terms of the dressed-states of the system \{atom+field\}.
%We point out that in the usual isotropic vacuum, $\text{Im}\left(\hat{\mathbf{G}}(\mathbf{r}_0,\mathbf{r}_0,\omega_0)\right)\propto \mathbb{I}$ ($\mathbb{I}$ is the unit tensor). So, one can see
%from Eq.~(\ref{ch4:eq:kappa}) that $\kappa_{12}$ vanishes because the transitions are orthogonal (\emph{i.e.} $\mathbf{d}_{01}^*\cdot\mathbf{d}_{02}=0$),
%and therefore there is no coherence $\rho_{12}$ [Eq.~(\ref{ch4:eq:coh_infty})].
%When $G_{xy}=0$,
Everything happens as if, after the emission of a photon ($t\rightarrow\infty$), the atom-field ``dressed-state'' is: 
\begin{equation}
  \begin{split}
  \ket{\psi(\infty)}=\frac{1}{\sqrt{d_{01}^2+d_{02}^2}}\frac{1}{\sqrt{\text{Im}(G_{xx})+\text{Im}(G_{yy})}}\\
  \times\left[ d_{01}\ket{1}\otimes \left(\sqrt{\text{Im}(G_{xx})}\ket{X}+\mathrm{i}\sqrt{\text{Im}(G_{yy})}\ket{Y}\right)\right.\\
    \left. +  d_{02}\ket{2}\otimes \left(\sqrt{\text{Im}(G_{xx})}\ket{X}-\mathrm{i}\sqrt{\text{Im}(G_{yy})}\ket{Y}\right)\right]\; ,
  \end{split}
\end{equation}
where $\ket{X}=1/\sqrt{2}(\ket{\sigma^+}+\ket{\sigma^-})$ [resp. $\ket{Y}=1/\sqrt{2}\mathrm{i}(\ket{\sigma^+}-\ket{\sigma^-})$] represents the state of photons emitted with a linear polarization along $\vec{x}$ [resp. $\vec{y}$].
Indeed, when tracing over the emitted photon, this fully agrees with Eq.~(\ref{ch4:rhocartesian}), and one also finds that $\rho_{ii}(\infty)=d_{0i}^2/(d_{01}^2+d_{02}^2)=\gamma_i/(\gamma_1+\gamma_2)$ (for $i=1,2$),
in agreement with Eq.~(\ref{ch4:eq:pop1}).

In isotropic vacuum, and when the two ground states are equally weighted ($d_{01}=d_{02}$, $\gamma_1=\gamma_2$),
it is well-known that the atom and the emitted photons are fully entangled \cite{scully1999quantum}: at the end of the decay process, the atom-field state is of the form:
\begin{equation}
\ket{\psi(\infty)}=\frac{1}{\sqrt{2}}\left(\ket{1}\otimes\ket{\sigma^+}+\ket{2}\otimes\ket{\sigma^-}\right)\; .
\end{equation}
The reduced state of each subsystem (atom and field as well) is thus fully incoherent which explains
why in isotropic vacuum we obtain a reduced density matrix $\rho(\infty)=\frac{1}{2} \mathbb{I}$.
It also explains why in order to observe quantum beats between the emitted photons in the vacuum a $V$-transition is necessary, and no quantum beats will appear in the case of a
$\Lambda$-transition (see Ref.~\cite{scully1999quantum}, Chapter 1.4).

However, if the back reaction of the environment fully eliminates the $X$-component of the polarization,
[\emph{i.e.} then $\text{Im}(G_{xx}) = 0$, which can be achieved with a metasurface as we will see later], the atom-field state at the end of the decay process is of the form:
\begin{equation}
\ket{\psi(\infty)}=\frac{1}{\sqrt{2}}\left(\ket{1}-\ket{2}\right)\otimes\frac{1}{\sqrt{2}\mathrm{i}}\left(\ket{\sigma^+}-\ket{\sigma^-}\right)\; .
\end{equation}
This atom-field state is factorisable (as it is the case for a $V$-transition in isotropic vacuum), but here the atom is in a coherent superposition of ground states
(whereas for a $V$-transition, it is the photon which is in a coherent superposition of two different modes).
This means that the reduced density matrix of each subsystem is a pure state.
In particular, we obtain here an atomic density matrix $\rho(\infty)$ equal to the 1-D projector:
$
\rho(\infty)=\frac{1}{2}\begin{bmatrix}
1 & -1\\
-1 & 1 \\
  \end{bmatrix}$.

The environment can thus act as a quantum eraser which erases the entanglement between
the atom and the field (emitted photon). According to the general complementary relation
between the entanglement of a system with its environment and the degree of coherence of the
reduced density matrix of this system \cite{jaeger1995two,englert1996fringe}, isotropic vacuum corresponds to the situation
where atom and field are maximally entangled so that their coherence is minimal (zero); on
the contrary, if the environment acts exactly as a polarization filter that destroys
linear polarization along $\vec{x}$, it also destroys the correlations (entanglement) between
the emitted photon and the two ground states, which fully restores the atomic coherence. In
realistic situations (as we will see in Section~\ref{ch4:sec:limitations}), only partial coherence is achieved, 
as an intermediary between these two
extreme cases (isotropic vacuum and ideal anisotropic vacuum).

\section{Metasurface designs}
\label{ch4:section:meta}

Vacuum anisotropy appears naturally in the near-field of a material media (see \emph{e.g.} \cite{PhysRevLett.103.063602}). For instance, anisotropic suppression of spontaneous emission of atoms located
between two close mirrors have been reported by W. Jhe \emph{et al.} \cite{jhe1987suppression}. Anisotropy of Casimir-Polder interactions between atoms and planar surfaces has also been investigated \cite{taillandier2014anisotropic} leading to atomic level mixing \cite{boustimi2001atom}.
Resonant nano-structures are also known to show important discrepancies between
$\text{Im}(G_{xx})$ and $\text{Im}(G_{yy})$ in the near-field of, for example, metallic nanodisks \cite{Thanopulos:19}, nanoparticles \cite{PhysRevA.83.055805,PhysRevB.95.075412} or graphene \cite{doi:10.1021/acs.jpcc.8b02703}.
Interestingly, although near-field interactions can dramatically enhance the QE spontaneous emission because of large $\text{Im}(G_{ii})$ values,
they are not better than far-field interactions for producing an optimum value of the anisotropy $A$ in Eq.~(\ref{ch4:rhocartesian}).

Metasurfaces acting as a spherical mirror with polarization-dependent responses
have been proposed to create anisotropic vacuum \cite{jha2015metasurface,jha2018spontaneous} in the far-field.
As a first example, looking at Eq.~(\ref{ch4:rhocartesian}), one can consider the ideal case of
a metasurface that perfectly reflects back to the QE half of its own emission only at a particular polarization, let say
the $x$-component, leading to perfect destructive interferences and thus $\text{Im}[G_{xx}(\bold{r}_0,\bold{r}_0,\omega_0)]=0$. Considering that the other polarization component (the $y$-component)
is not affected and thus $\text{Im}[G_{yy}(\bold{r}_0,\bold{r}_0,\omega_0)]=\gamma_0/2$, its value in vacuum, such a metasurface might lead to an optimum anisotropy $A=-1$.
This was the strategy followed in Ref.~\cite{jha2015metasurface} in order to induce the coherent population transfer predicted in Ref.~\cite{agarwal2000anisotropic} for a $V$-configuration.

A metasurface can alternatively be designed as acting on circular polarizations.
To clarify this, instead of expressing the quantities appearing in Eqs.~(\ref{ch4:eq:gamma_Green}) and (\ref{ch4:eq:kappa})
in Cartesian coordinates [as done to obtain  Eq.~(\ref{ch4:rhocartesian})], let us express the Green tensor and dipole moments in the spherical basis
$(\vec{\varepsilon}_+,\vec{\varepsilon}_-,\vec{\varepsilon_0})$, where $\vec{\varepsilon}_\pm=(\vec{x}\pm i\vec{y})/\sqrt{2}$ and $\vec{\varepsilon_0}=\vec{z}$.
Then, by plugging these expressions into Eq.~(\ref{ch4:eq:coh_infty}), one finds the following expression for the coherence:
\begin{equation}
  \rho_{12}(\infty)=\underbrace{\frac{d_{01}d_{02}}{d_{01}^2+d_{02}^2}}_R\times
  \underbrace{\frac{\text{Im}(G_{+-})}{\text{Im}(G_{++})}}_A\; ,
  \label{ch4:rhocylindrical}
\end{equation}
where the Green tensor spherical components have again to be evaluated at the position of the quantum emitter $\bold{r}_0$ and at the transition frequency $\omega_0$.
This expression is equivalent to Eq.~(\ref{ch4:rhocartesian}), since the Green tensor components in the different basis verify the following relations: $G_{+-}=\frac{1}{2}\left(G_{xx}-G_{yy}-\mathrm{i}2G_{xy}\right)$
and $G_{++}=G_{--}=\frac{1}{2}(G_{xx}+G_{yy})$. The form of Eq.~(\ref{ch4:rhocylindrical}) suggests that a metasurface that mixes the circular polarizations $\sigma^+$ and $\sigma^-$, and thus
leading to a non-null cross-term $\text{Im}[G_{+-}(\bold{r}_0,\bold{r}_0,\omega_0)]\neq 0$, might create a coherence. Ideally, if the metasurface totally inverses the absolute rotation direction
of the electric field with respect to that of the incident
circularly polarized one, one will have $\text{Im}[G_{+-}(\bold{r}_0,\bold{r}_0,\omega_0)]=\text{Im}[G_{++}(\bold{r}_0,\bold{r}_0,\omega_0)]$, and thus a maximum anisotropy $A=1$.
This strategy was employed in Ref.~\cite{jha2018spontaneous} in order to induce a coherence between the two excited states in a $V$-configuration.

In this Part, we present the two designs of metasurface discussed in the above examples, we compare their performances,
and we assess the value of the induced coherence, taking into account the limitations of such designs.
But first of all, we present in Section~\ref{sect:pma} the general approach used to make the designs.

\subsection{Phase-mapping approach}
\label{sect:pma}

The problem considered here is the interaction between a planar (meta)surface and an electric dipole source of emission wavelength
$\lambda_0$ located at a distance $d$ above the surface. For an emitter located at remote distances (in the far-field $d\gg\lambda_0$), the interaction
will be efficient only if the metasurface is able to reflect and
focus back the light originating from the ``point'' dipole source. Thus, the metasurface must be optically equivalent to a spherical mirror of focal length $f=d/2$, by producing the following spherical phase profile:
%This already sets the experimental constraints on the QE: one must be able to position the emitter with a precision of $\lambda$ at remote distances $d\gg\lambda$.
\begin{equation}
  \varphi(\mathbf{r})= \pi-2k_0|\mathbf{r}-\mathbf{r}_0|\; (\mathrm{mod}\,2\pi)\; ,
  \label{ch4:eq:phase_profile}
  \end{equation}
where $k_0=2\pi/\lambda_0$, $\mathbf{r}$ are the coordinates of the points of the metasurface, and $\mathbf{r}_0$ are the coordinates of the QE.

\begin{figure}[h!]
   \centering
   \includegraphics[scale=0.5]{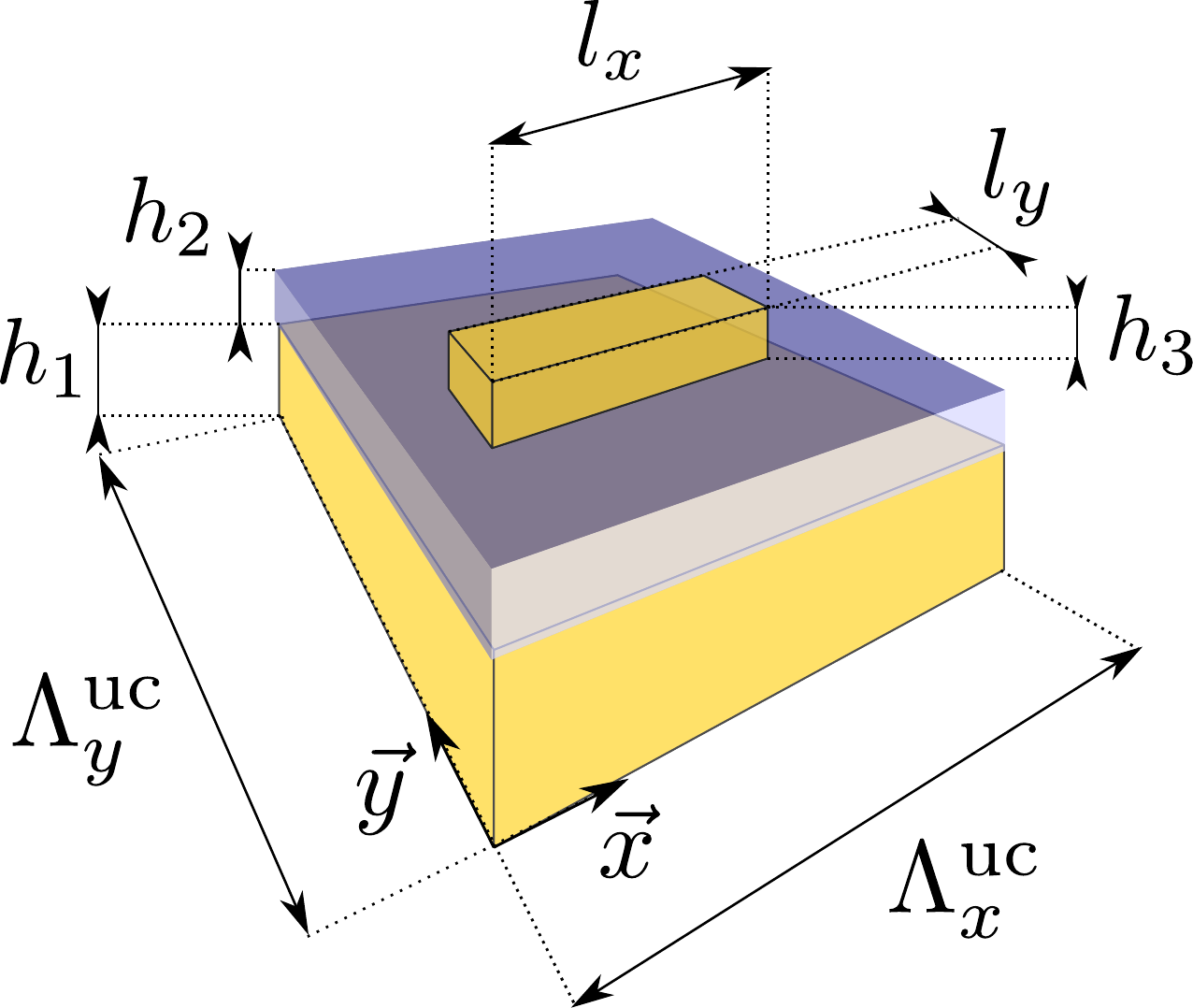}
   \caption{Unit-cell of a reflect-array metasurface made of: a metallic mirror of thickness $h_1$, a dielectric spacer of thickness $h_2$, and a rectangular nanoantenna of dimensions $l_x\times l_y$
     and of thickness $h_3$. The dimensions of the unit-cell are: $\Lambda_x^\text{uc}\times\Lambda_y^\text{uc}$.}
   \label{ch4:fig:unitcell}
\end{figure}

We parametrize the problem as follows: the points $\mathbf{r}$ lie
in the plane $z=0$: $\mathbf{r}=(x,y,0)$, and $\mathbf{r}_0=(0,0,d)$.
In other words, the phase accumulated through propagation should be compensated in each point $\mathbf{r}$ of the flat metasurface --- hence the minus sign in Eq.~(\ref{ch4:eq:phase_profile}) --- by a phase-shift
corresponding to the phase profile given in Eq.~(\ref{ch4:eq:phase_profile}).
Such metasurfaces create interferences and a diffraction limited spot (at the position of the QE $\mathbf{r}_0$), and are the equivalent in reflection of \emph{metalenses} \cite{genevet2017recent,lalanne2017metalenses}.
They can be implemented using metallic subwavelength reflect-arrays, made of a metallic mirror, a dielectric spacer and subwavelength structures (also called \emph{meta-atoms} or nanoantennas)
patterned on top (see Fig.~\ref{ch4:fig:unitcell}). 

By carefully choosing and positionning the meta-atoms, the metasurface can induce local phase-shifts that mimic the spherical phase profile given by Eq.~(\ref{ch4:eq:phase_profile}).
This is the principle of the \emph{phase-mapping approach}.
Obviously, each design is specific for a couple of parameters $\{\lambda_0,d\}$, so a modification of one of these parameters leads to a new design.
Good power reflectances were reported for such metasurfaces at normal incidence: about $80\%$ for gold reflectarrays in the range $700-1100\,\text{nm}$ \cite{sun2012high,pors2013gap,zheng2015metasurface,jha2015metasurface}
and up to $90\%$ for silver reflectarrays around $640-670\,\text{nm}$ \cite{jha2017metasurface,jha2018spontaneous}.

%The performances and properties of a metasurface depends on the microstructure, that is how the subwavelength elements, also called meta-atoms, are patterned.
In order to design the metasurface, one usually extrapolates its properties from the computations of an \emph{infinite periodic grating}.
Such an approach assumes that \emph{locally}, the properties of the metasurface are close to the one of a periodic grating, which is valid if the meta-atoms behave independently
\cite{lalanne2017metalenses}, and is refered to as \emph{local-periodicity approach}.
Adopting this approach, all the numerical simulations in this paper are done using the open-source code RETICOLO software for grating analysis \cite{hugonin2005reticolo},
%developped by J.P. Hugonin and P. Lalanne, Institut d'Optique, Palaiseau, France (2005),
which implements a frequency-domain modal method known as the Rigorous Coupled Wave Analysis (RCWA) \cite{moharam1995formulation,li1997new,lalanne1998computation,popov2000grating}.
Since it is a Fourier modal method, we specify for each simulation the number of Fourier modes retained for the computation (which are given for the direction $\vec{x}$,
and we use the same number of modes for the direction $\vec{y}$).
Moreover, all the meta-atoms considered here are nanorods that respect a mirror symmetry, and therefore throughout this paper: $G_{xy}=0$ \cite{luo2015photonic}.

In the next Sections~\ref{sec:resonant-phase-delay} and \ref{sec:geometa}, we present two designs aiming at creating the coherence in the QE, and we characterize their performances.

%Employing Eq.~(\ref{ch4:int}), the criterion (i) is satisfied when:
%\begin{equation}
  %G_{zz}^0(\mathbf{r}_m,\mathbf{r}_0)R_{++}(\mathbf{r}_m)G_{zz}^0(\mathbf{r}_0,\mathbf{r}_m)=0 \Rightarrow R_{++}(\mathbf{r}_m)=0
  %\label{ch4:criterion_i_bis}
%\end{equation}

\subsection{Design based on resonant-phase delays}
\label{sec:resonant-phase-delay}

In this Section, we design the metasurface discussed in the first example above --- inspired from Eq.~(\ref{ch4:rhocartesian}) --- that must have the following optical properties:
%As seen from Eq.~(\ref{ch4:rhocartesian}), a way to reach the maximum value for the visibility $V$, and thus for coherence $\rho_{12}$, is to cancel one of the two components $\text{Im}[G_{xx}(\bold{r}_0,\bold{r}_0,\omega_0)]$
%or $\text{Im}[G_{yy}(\bold{r}_0,\bold{r}_0,\omega_0)]$.
%This strategy was followed in \cite{jha2015metasurface},
%Such a remote effect was already predicted with the use of mirrors \cite{yang2008quantum,hetet2010qed}.
%where the authors proposed a metasurface with the following optical properties:
(i) The metasurface acts as a spherical mirror \emph{only} for a linearly-polarized light along $\vec{x}$, resulting in $G_{xx}(\bold{r}_0,\bold{r}_0,\omega_0)=0$ (destructive interferences);
(ii) The metasurface acts as a planar mirror for a linearly-polarized light along $\vec{y}$, so $G_{yy}(\bold{r}_0,\bold{r}_0,\omega_0)$ is untouched.
%We recall that the plans $(x,y)$ are parallel to the plan metasurface, which is the chosen as the particular plan lying in $z=0$.
Such a metasurface can be built from anisotropic resonant nanoantennas,
using for example metallic nanorods (like the one represented in Fig.~\ref{ch4:fig:unitcell}) with: Varying lengths along $\vec{x}$, in order to tune the resonance and to
induce different phase-shifts or \emph{resonant-phase delays} on a $x$-polarized light that reproduce the spherical phase profile of Eq.~(\ref{ch4:eq:phase_profile});
And the same width along $\vec{y}$, in order to induce a constant phase-shift on a $y$-polarized light that produces a flat phase profile \cite{sun2012high,pors2013gap,jha2015metasurface}.

\begin{figure}[h!]
   \centering
   \includegraphics[width=\linewidth]{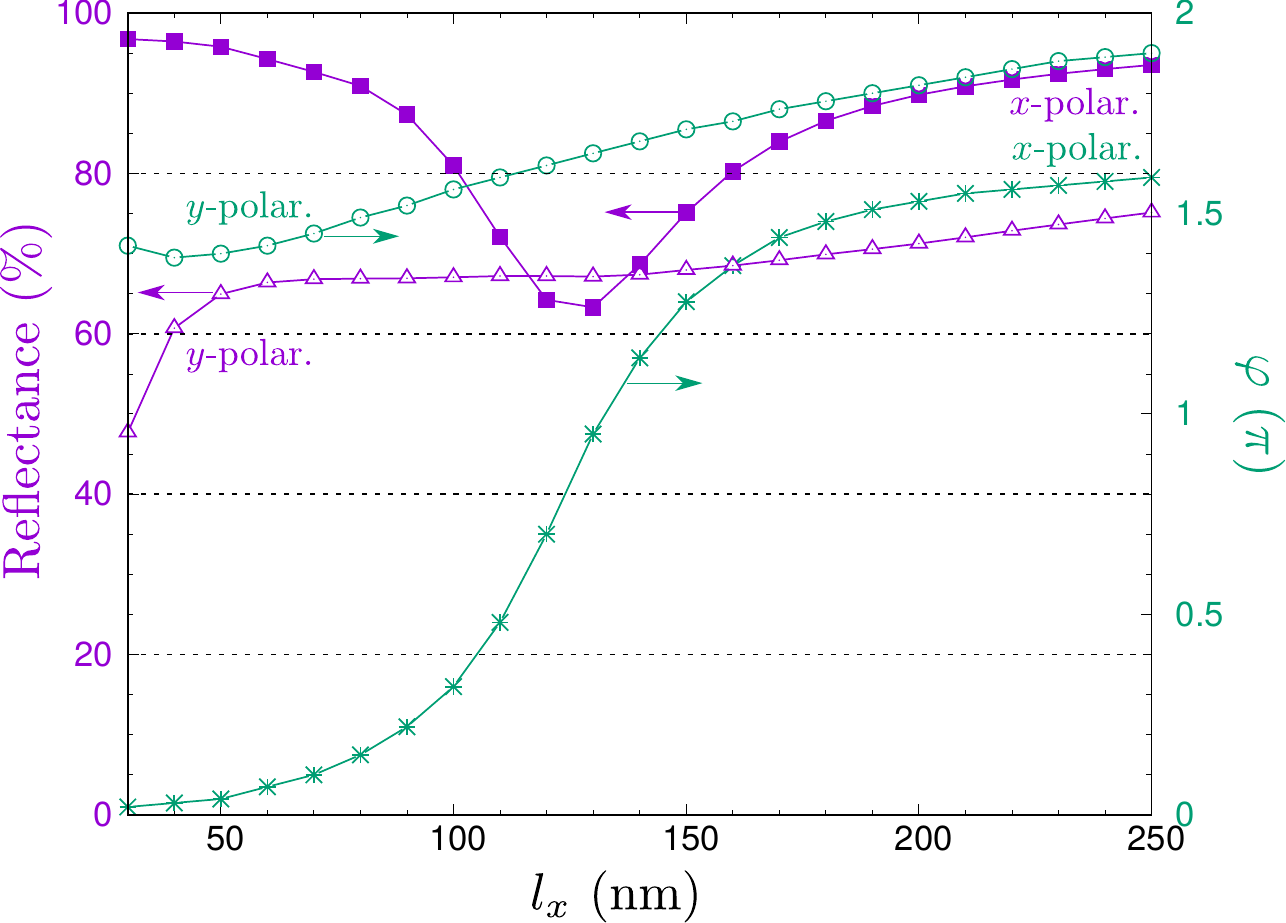}
   \caption{Power reflectance (in purple) and phase-shift $\varphi$ (in green) of an incident $x$-polarized [resp. $y$-polarized] wave as a function of the length $l_x$
     of the nanoantennas, computed for a 2-D grating (see main text). The symbols represent the simulated points, and the solid lines are guide-to-the-eyes. The dotted black lines are spaced by $2\pi/5$.
   The number of Fourier modes used for the simulations is 30.}
   \label{ch4:resonant-phase}
\end{figure}

For the simulations, we consider a 2-D grating made of unit-cells of the type presented in Fig.~\ref{ch4:fig:unitcell} with lateral dimensions of
$\Lambda_x^\text{uc}\times\Lambda_y^\text{uc}=300\,\text{nm}\times 150\,\text{nm}$,
and made of a gold mirror and a dielectric film of SiO$_2$ with respective thicknesses $h_1=130\,\text{nm}$ and $h_2=50\,\text{nm}$, and a gold
nanorod patterned on top with fixed width $l_y=100\,\text{nm}$  and thickness $h_3=30\,\text{nm}$.
The wavelength is chosen at $\lambda_0=852\,$nm, which corresponds to the D2-line of cesium atom.
At this wavelength, the refractive indices are $n=0.16 + \mathrm{i}5.34$ for gold and $n=1.45$ for SiO$_2$.
In Fig.~\ref{ch4:resonant-phase}, we computed the phase-shifts (in green) and the efficiencies in reflection (in purple) of such a 2-D grating,
for incident $x$ and $y$-polarized waves at normal incidence, as a function of the length $l_x$ of the nanorod.
One can see that the phase-shift induced on a $x$-polarized wave (green crosses) spans over $8\pi/5$ ($1.6\pi$), which corresponds to 4/5 of the $2\pi$ phase space,
while the phase-shift induced on a $y$-polarized wave (green circles) is rather flat. 
We can therefore choose five nanoantennas to sample the entire phase space of $2\pi$, with respective phase-shifts of: $0$, $2\pi/5$, $4\pi/5$, $6\pi/5$ and $8\pi/5$
(intersection with the dotted black lines spaced by $2\pi/5$, see dimensions in Table~\ref{ch4:table:nanoantennas}).
Moreover, the power reflectance of the $x$-polarized wave (purple squares), which is the only one that matters, is relatively good, remaining between $63\%$ and $97\%$, the losses being due to
absorption by the metal (we check that the gold mirror is thick enough and that there is no transmission losses).

\begin{table}[htbp]
  \centering
  \def\arraystretch{1.25}
\begin{tabular}{ccc}
  \hline
  \hline
\multicolumn{1}{c}{$\;\;$ nanoantenna $\;\;$} &
\multicolumn{1}{c}{$\;\;\;\;\;\;$ $l_x$ $\;\;\;\;\;\;$} &
\multicolumn{1}{c}{$\;\;\;\;\;\;$ $l_y$ $\;\;\;\;\;\;$} \\
\hline
\#$1$ & $30\,\text{nm}$ & $100\,\text{nm}$ \\

\#$2$ & $105\,\text{nm}$ & $100\,\text{nm}$ \\

\#$3$ & $125\,\text{nm}$ & $100\,\text{nm}$ \\

\#$4$ & $145\,\text{nm}$ & $100\,\text{nm}$ \\

\#$5$ & $250\,\text{nm}$ & $100\,\text{nm}$ \\
\hline
\hline
\end{tabular}
\caption{Nanoantennas dimensions for sampling the phase-space from 0 to $2\pi$.}
\label{ch4:table:nanoantennas}
\end{table}

\begin{figure}[b]
   \centering
   \includegraphics[width=\linewidth]{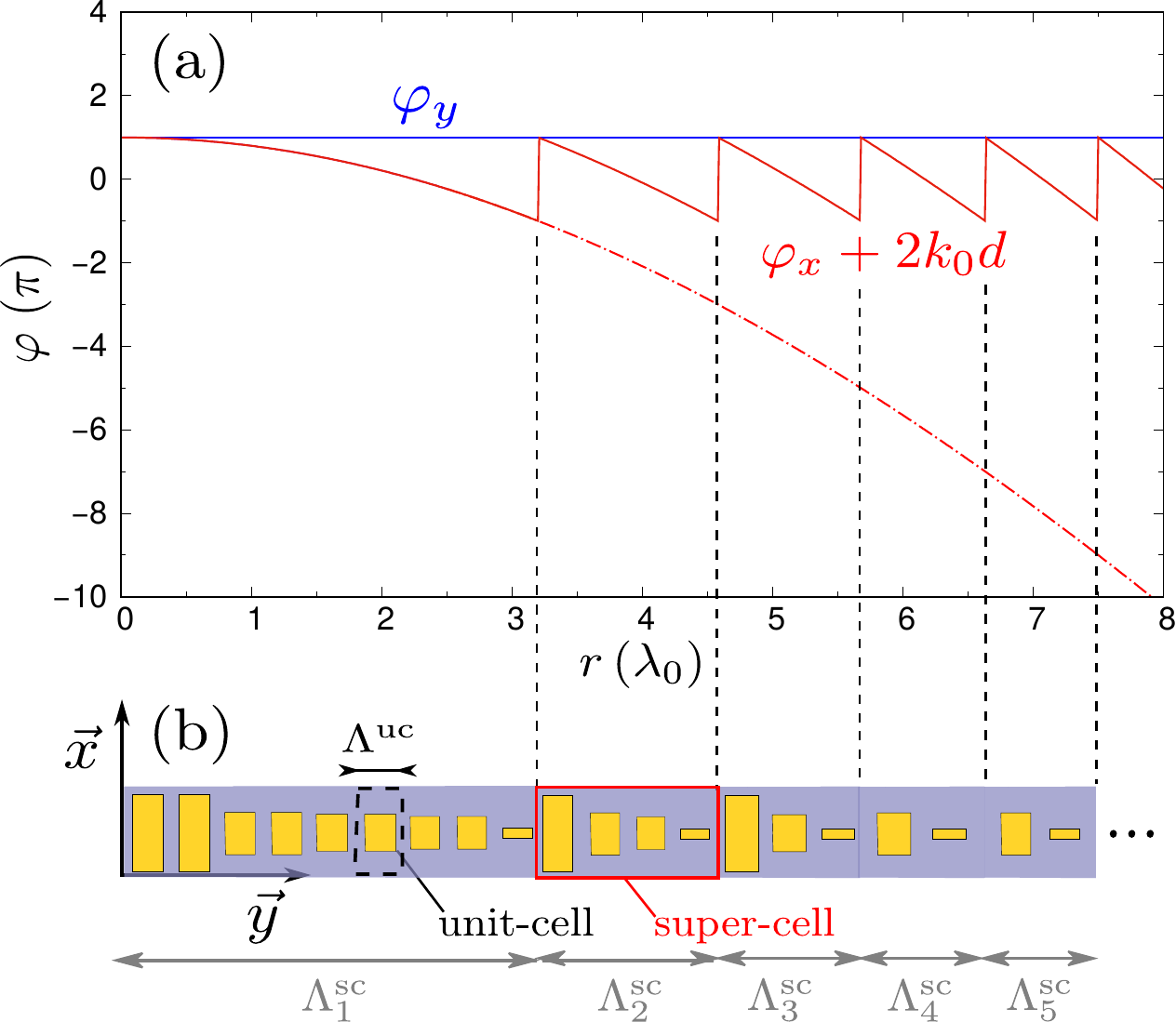}
   \caption{Illustration of the phase-mapping approach for the 1-D design of the resonant-phase delay metasurface. (a) Phase profiles to be encoded by the metasurface: the wrapped [resp. unwrapped] spherical phase profile $\varphi_x$
     of Eq.~(\ref{ch4:eq:phase_profile}) (red full line) [resp. red dashed line] desired for the $x$-polarization, and the flat phase profile $\varphi_y$ (blue line) desired for the $y$-polarization,
starting from the center of the metasurface at $r=0$. (b) Corresponding nanoantennas to encode the desired phase-shifts $\varphi_x$ and $\varphi_y$.
     The unit-cells of length $\Lambda^\text{uc}$ (black dashed box) containing the nanoantennas are encompassed into super-cells of length $\Lambda^\text{sc}$ (red box),
     spanning the $2\pi$ phase-space.}
   \label{ch4:phase-mapping}
\end{figure}

The design of the metasurface is achieved after combining these five nanoantennas while employing the following rules: All nanoantennas must be parallel (the varying length $l_x$ always oriented along the $\vec{x}$ axis),
and patterned after Eq.~(\ref{ch4:eq:phase_profile}) according to the phase-mapping approach.
This design is illustrated in ``1-D'' in Fig.~\ref{ch4:phase-mapping}: In Fig.~\ref{ch4:phase-mapping} (a), we plot the ideal unwrapped (resp. wrapped) phase profile of Eq.~(\ref{ch4:eq:phase_profile})
of a $x$-polarized wave in dashed red (resp. full red), and the ideal flat phase profile of a $y$-polarized wave in blue, starting from the center of the metasurface at $r=0$;
In Fig.~\ref{ch4:phase-mapping} (a), we represent a slice of the metasurface where the nanoantennas are distributed into super-cells (one of them is highlighted in the red box)
that sample the $2\pi$ phase-space, mimicking the phase profiles of Fig.~\ref{ch4:phase-mapping} (a). The size of the super-cells is maximum at the center of the metasurface,
and progressively decreases with distance from the center because the spherical phase
profile varies more rapidly.

In addition to the absorption losses, the sampling of the phase by \emph{discrete} elements in the phase-mapping approach also limits the performances of the metasurface (one talks about \emph{discretization losses}).
In order to assess these discretization losses, we compute the performances in the canonical case of a \emph{linear-phase gradient metasurface} \cite{sun2012high,pors2013gap},
which behaves as a \emph{blazed grating} that diffracts entirely into the diffraction order $m=-1$ only for an incident $x$-polarized wave.
Such a gradient metasurface is made of a same super-cell containing nanoantennas that sample the phase regularly from $2\pi$ to $0$ (and from $0$ to $2\pi$ to diffract into the order $m=+1$), repeated with periodic boundary conditions.

\begin{figure}[b]
   \centering
   \includegraphics[width=\linewidth]{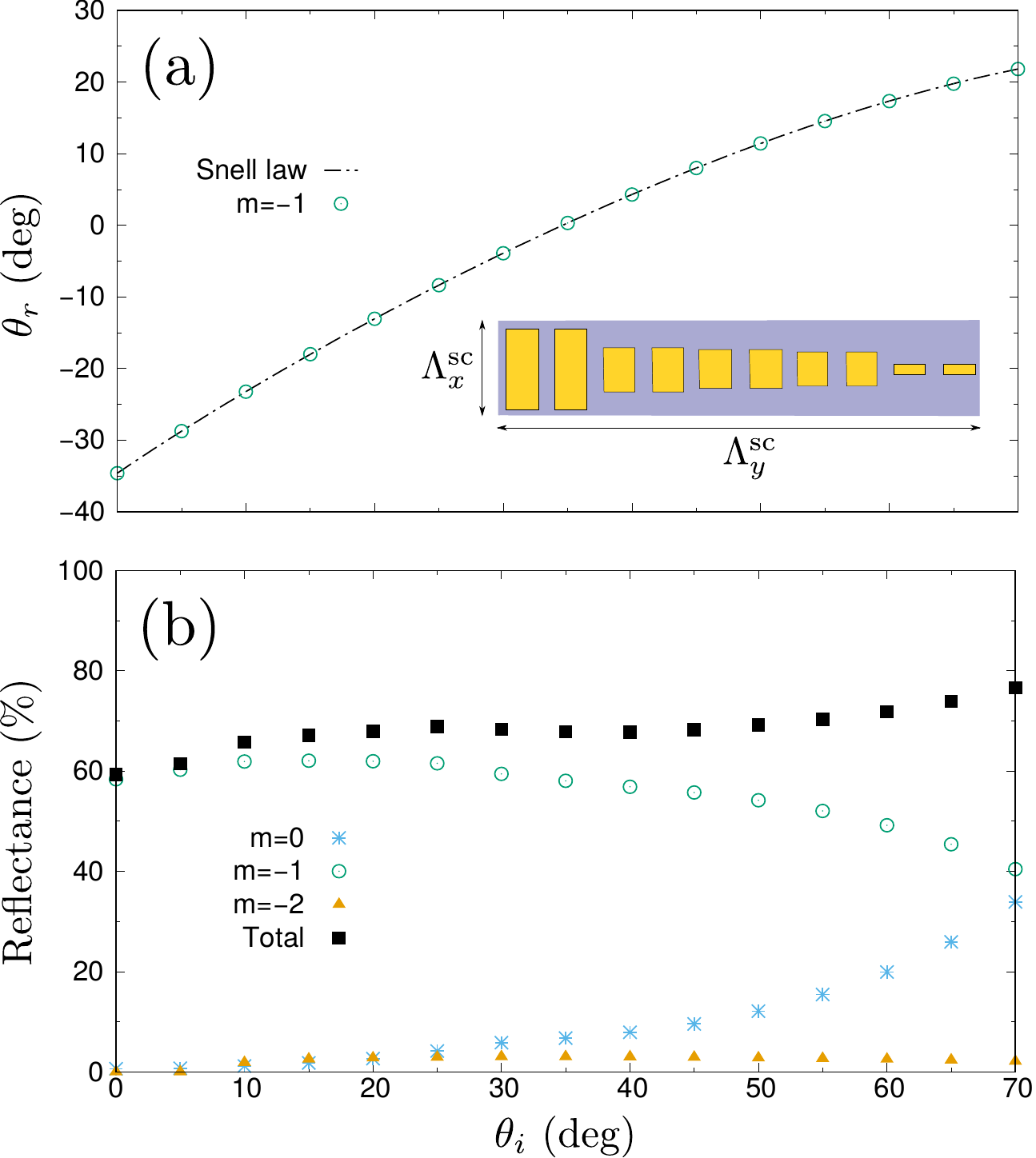}
   \caption{Diffraction performances of a linear-phase gradient metasurface.
     The inset shows a super-cell of the gradient metasurface of size $\Lambda_x^\text{sc}\times \Lambda_y^\text{sc}$ (see main text). (a) Reflection angle $\theta_r$ in the diffracted order $m=-1$ of an incident plane wave polarized along $\vec{x}$ as a function of the incident angle $\theta_i$ (green circles).
   The generalized Snell's law of reflection [Eq.~(\ref{eq:snell})] is also plotted (black dashed line).
   (b) Power reflectance in the diffracted orders $m=0,-1,-2$ (blue stars, green circles, and orange triangles, respectively) and total reflection efficiency (black squares)
   of an incident plane wave polarized along $\vec{x}$ as a function of the incident angle
   $\theta_i$. The number of Fourier modes used to compute the reflectance is 30.}
      \label{ch4:angle_supercell}
\end{figure}

For the simulations, we consider a linear-phase gradient metasurface made of super-cells of dimensions $\Lambda_x^\text{sc}\times \Lambda_y^\text{sc}=300\,$nm$\times 1500\,$nm, in which the five nanoantennas previously selected are embedded
into unit-cells, with the same dimensions as previously, and repeated twice [see inset in Fig.~\ref{ch4:angle_supercell} (a)]. The working wavelength is still $852\,\text{nm}$, as previously.
The angle of the diffracted order $m=-1$ (reflection angle $\theta_r$) is given in terms of the angle of the incident wave (incident angle $\theta_i$) by the generalized Snell's law of reflection \cite{genevet2017recent}:
\begin{equation}
  \text{sin}\left(\theta_r\right)=\text{sin}\left(\theta_i\right)+\frac{\lambda_0}{2\pi}\frac{\partial \varphi}{\partial y}\; ,
  \label{eq:snell}
\end{equation}
where in our case $\partial\varphi/\partial y=-2\pi/\Lambda_y^\text{sc}$ with $\Lambda_y^\text{sc}=1500\,$nm.
We check that we perfectly recover this law in Fig.~\ref{ch4:angle_supercell} (a) for an incident $x$-polarized wave.
Thus, one can see that the diffraction angle is the same either for a periodic blazed grating or for a smooth linear-gradient metasurface, because it only depends on the period
and not on the underlying structure \cite{Larouche:12}.

In Fig.~\ref{ch4:angle_supercell} (b), we computed the power reflectance of the diffracted order $m=-1$ for an incident $x$-polarized wave (green circles) as a function of the incident angle $\theta_i$,
and compare it with other dominant orders $m=0$ (blue stars) and $m=-2$ (orange triangles). The total power reflectance is also shown (dark squares).
Firstly, the total power reflectance, which varies between $59\%$ and $77\%$, reveals absorption losses between $23\%$ and $41\%$, depending on the incident angle.
Secondly, one can see that the reflectance of the order $m=-1$ is about $60\%$ for incident angles $\theta_i$ up to $30\degree$,
and then decreases until $40\%$ for an incident angle of $70\degree$, while mostly the reflectance of the order $m=0$ increases.
This reveals that while the reflectance into a given order depends on the incident angle $\theta_i$, it is relatively robust with the variations of $\theta_i$
(1/3 decrease of the reflectance of the order $m=-1$ over $70\degree$).

%In Fig.~\ref{ch4:angle_supercell} (a), we show the reflectance properties as a metasurface mimicking a linear-gradient of the phase.
The final metasurface is more complex than a linear-phase gradient metasurface since it is made of super-cells of different sizes.
One can show from Eq.~(\ref{ch4:eq:phase_profile}) that the largest super-cell starts at $r=0$ (at the center of the metasurface) and has a length of $\Lambda_\text{max}^{\text{sc}}=\sqrt{d\lambda_0}$,
and that the length of the next super-cells quickly converges towards the minimum length of $\Lambda_\text{min}^{\text{sc}}=\lambda_0/2$.
%(see Appendix~\ref{ch5:app:sc} for more detail).
In Table~\ref{ch4:tab:supercell}, for a design working at $\{\lambda_0=852\,\text{nm},d=10\lambda_0\}$, we give:
the length $\Lambda^{\text{sc}}$ of the first five super-cells represented in Fig.~\ref{ch4:phase-mapping} (b) (and labelled $n=1,...,5$ starting from the center);
the number of unit-cells $N$ per super-cell, considering a unit-cell of fixed length $\Lambda^\text{uc}=300\,\text{nm}$ ($\sim 0.35 \lambda_0$).
One can see that the number of unit-cells --- and therefore of nanoantennas --- quickly drops from $9$ (first super-cell) to $2$ (forth super-cell).
Consequently, the sampling of the phase deteriorates, leading to higher discretization losses.

We computed in Table~\ref{ch4:tab:supercell} the power reflectance of the order $m=-1$ (for an incident $x$-polarized wave) for different linear-phase gradient metasurfaces made of these super-cells,
and taking into account the incident angle $\theta_i$ (also shown) at which the light impinges the super-cell in the final metasurface.
One can see that the reflectance decreases as the number of unit-cells per super-cell decreases; in other words, the discretization losses increase.

In summary, the performances of the metasurface are reduced for two main reasons: the absorption losses and the discretization losses due to the finite number of
unit-cells used to sample the phase; They are better in the center of the metasurface, and deteriorate quickly when getting further from the center (or increasing of the incident angle),
which limits the numerical aperture (NA) of the metasurface.

%Let us now assess the efficiency of the metasurface made of the addition of these supercells and appearing in Eq.~(\ref{sec:discuss:gamma})
%in the form of the coefficient $R$ that we will use to calculate the coherence of Eq.~(\ref{ch4:rhocartesianrecallnew}).
%To a metasurface made of the $n$ first supercells corresponds the numerical aperture: $\text{NA}(n)=\sum_{k=1}^n\Lambda_k^{\text{sc}}/\sqrt{[\sum_{k=1}^n\Lambda_k^{\text{sc}}]^2+d^2}$.
%We assess the efficiency of a metasurface made of $n$ supercells from the efficiencies calculated in Table~\ref{ch4:tab:supercell} using the following formula:
%\begin{equation}
%  R_n=\sum_{k=1}^{n}R_k\times\frac{\Omega_k-\Omega_{k-1}}{\Omega_n}
%  \label{discuussRR}
%\end{equation}
%where $R_n$ and $\Omega_n=2\pi\left[1-\sqrt{1-\text{NA}(n)^2}\right]$ are respectively the efficiency and the solid angle associated to the metasurface made of the first $n$ supercells;
%$R_k$ is the efficiency the supercell $k$ given in Table~\ref{ch4:tab:supercell} and $\Omega_k=2\pi\left[1-\sqrt{1-\text{NA}(k)^2}\right]$ is the solid angle associated to the metasurface made of the first $k$ supercells
%(and we denote $\Omega_0\equiv 0$).
%We plotted this quantity as a function of the numerical aperture of the metasurface in Fig.~\ref{ch4:lastfig} (circles).
%Note that even if this efficiency was calculated for the particular configuration $\lambda_0=852\,\text{nm}$ and $d=10\lambda_0$, it can be shown that
%it is invariant with the ratio $\lambda_0/d$ (see Appendix~\ref{ch5:app:uc} for a discussion).

\begin{center}
\begin{table} [h!]
\begin{center}
\def\arraystretch{1.25}
\begin{tabular}{ccccc}
\hline
\hline
\multicolumn{1}{c}{Super-cell n} &
\multicolumn{1}{c}{$\;\;$ $\Lambda^{\text{sc}}\,(\lambda_0)$ $\;\;$} &
\multicolumn{1}{c}{$\;\;\;\;$ $N$ $\;\;\;\;$} &
\multicolumn{1}{c}{$\;\;$ $\theta_i\,$($\degree$)} &
\multicolumn{1}{c}{$\;\;$ Reflectance$\,(\%)$ $\;\;$}\\
%\multicolumn{1}{c}{} &
%\multicolumn{1}{c}{dipole} &
%\multicolumn{1}{c}{quadrupole} &
%\multicolumn{1}{c}{octupole} \\
\hline
$1$ & $3.17$ & $9$ & $0$ &$60$\\
$2$ & $1.41$ & $4$ & $17.6$ & $55$\\
$3$ & $1.06$ & $3$ & $24.6$ & $50$\\
$4$ & $0.94$ & $2$ & $29.4$ & $30$\\
$5$ & $0.82$ & $2$ & $33.3$ & $30$\\
$\infty$ & $ 0.50 $ & $1$ & $90.0$ & $0$\\
\hline
\hline
\end{tabular}
\end{center}
\caption{Characteristics of the super-cells of the metasurface shown in Fig.~\ref{ch4:phase-mapping}, labelled by integer $n=1,2,3,4,5$ (starting from the center of the metasurface): length $\Lambda^{\text{sc}}$ (in units of $\lambda_0$),
  number of unit-cells $N$ per super-cell, incident angle $\theta_i$ of the light impinging the super-cell, and power reflectance in the order $m=-1$ (computed for a linear-phase gradient metasurface made of the super-cell and at the incident angle $\theta_i$).
The number of Fourier modes used to compute the reflectance is 30.}
\label{ch4:tab:supercell}
\end{table}
\end{center}

\subsection{Design based on geometric phases}
\label{sec:geometa}

\begin{figure}[t]
   \centering
   \includegraphics[width=\linewidth]{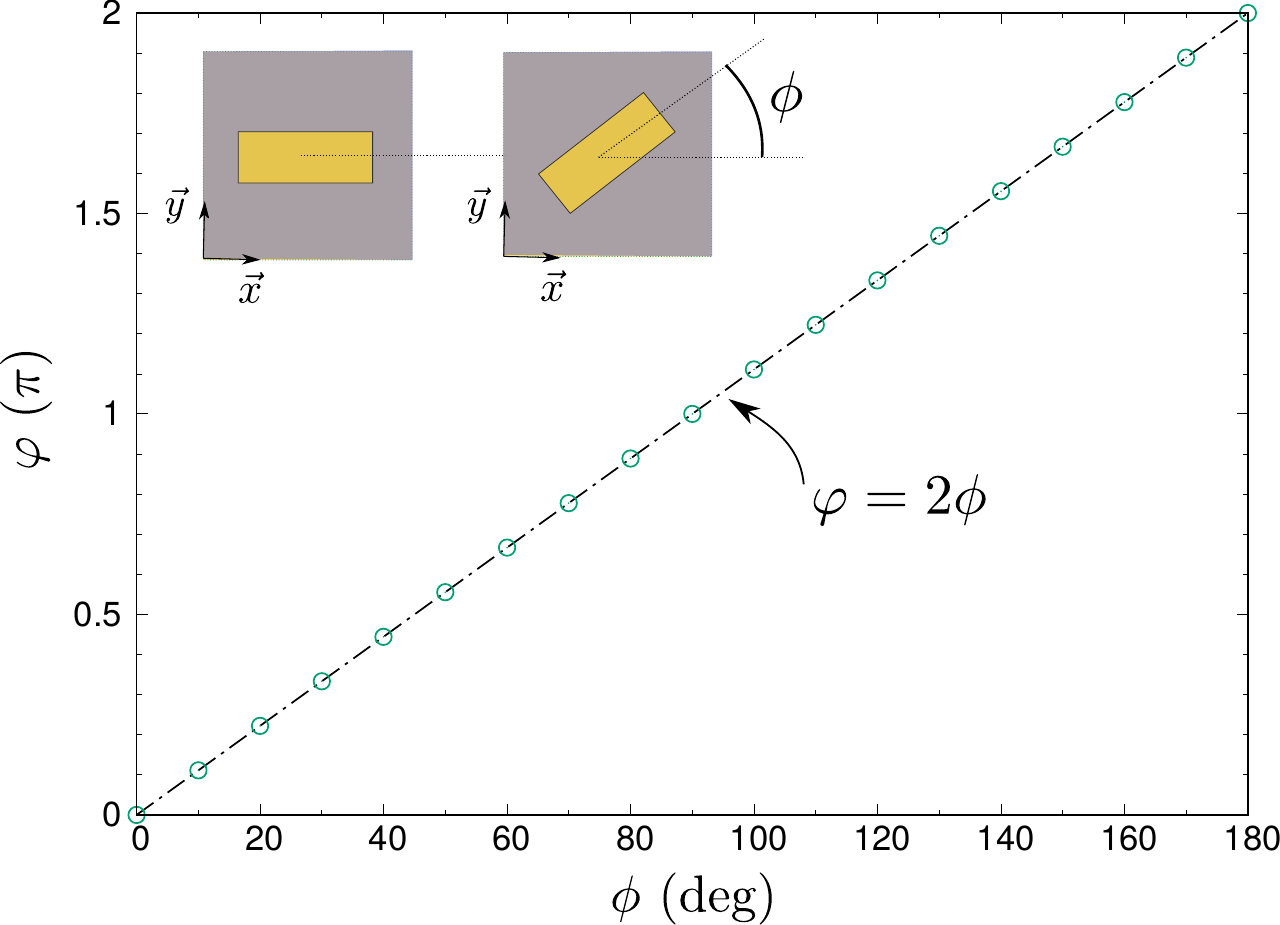}
   \caption{Geometric phase $\varphi$ as a function of the rotation angle $\phi$ of the nanorods in the plane $(\vec{x},\vec{y})$ (see inset),
     computed for a 2-D grating (see main text) (green circles). The analytical expression [Eq.~(\ref{eq:geom_phase})] is also plotted (black dashed line).
   The number of Fourier modes used to compute the geometric phase is 30.}
   \label{ch4:phase}
\end{figure}

In this Section, we design the metasurface discussed in the second example above --- inspired from Eq.~(\ref{ch4:rhocylindrical}) --- that must have the following properties:
(i) The metasurface acts as a spherical mirror;
(ii) Upon reflection, the metasurface totally inverses the absolute rotation direction of the electric field with respect to that of the incident
circularly polarized one.
This inversion of the electric field rotation can be achieved by using nanoantennas which act as half wave plates, as the result of a phase
delay of $\pi$ between the long and short axes of the nanoantennas \cite{luo2015photonic,zheng2015metasurface}.
Moreover, a phase-shift, called \emph{geometric phase} or \emph{Pancharatnam–Berry phase}, which depends on the orientation of the antenna, is acquired through this inversion, according to \cite{pancharatnam1956generalized,berry1987adiabatic}:
\begin{equation}
  \varphi=2\phi\; ,
  \label{eq:geom_phase}
\end{equation}  
where $\phi$ denotes the angle by which the antenna is rotated (see inset in Fig.~\ref{ch4:phase}).
This phase-shift is of geometric origin since it is solely due to the orientation of the nanoantenna and not to its resonance properties.
Thus, the spherical phase profile can be built by mapping the orientation of the nanoantennas [Eq.~(\ref{eq:geom_phase})] into the spherical phase profile [Eq.~(\ref{ch4:eq:phase_profile})].

For the simulations, we consider a 2-D grating made of unit-cells of the type presented in Fig.~\ref{ch4:fig:unitcell}
with lateral dimensions of $\Lambda_x^\text{uc}\times\Lambda_y^\text{uc}=300\,\text{nm}\times 300\,\text{nm}$,
and made of a gold mirror and a dielectric film of MgF$_2$ with respective thicknesses $h_1=130\,\text{nm}$ and $h_2=90\,\text{nm}$, and a gold
nanorod patterned on top lateral dimensions $l_x=200\,$nm and $l_y=80\,$nm and thickness $h_3=30\,\text{nm}$, following Refs.~\cite{luo2015photonic,zheng2015metasurface}.
The working wavelength is $852\,$nm, and the refractive indices are $n=0.16 + \mathrm{i}5.34$ for gold and $n=1.37$ for MgF$_2$.
For such a system, the phase-shift for a light polarized along $\vec{x}$ and a light polarized along $\vec{y}$ is $\pi$ upon reflection, at $852\,$nm.
Thus, the system acts as a half-wave plate working in reflection.

\begin{figure}%[h!]
   \centering
   \includegraphics[width=\linewidth]{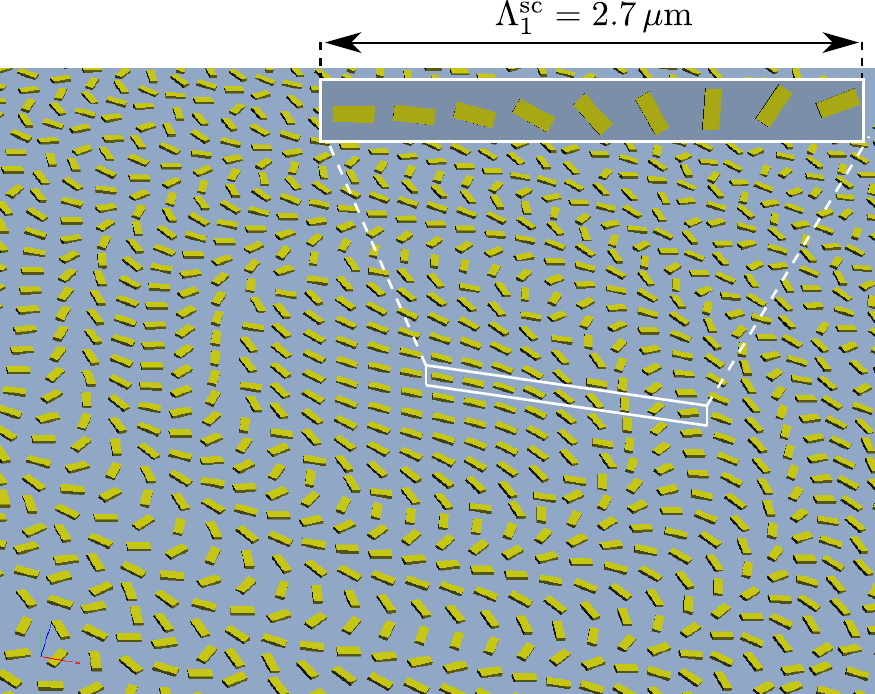}
   \caption{3-D design of the geometric metasurface, made for a distance of the dipole source $d=10\lambda_0$ from the metasurface, and an emission wavelength of $\lambda_0=852\,\text{nm}$.
   The white box highlights the first super-cell (starting from the center) of size $\Lambda_1^\text{sc}=2.7\,\mu\text{m}$, made of nine nanorods. This computer aided design was drawn using the software SOLIDWORKS developped by Dassault Syst\`emes.}
   \label{ch4:metasurfacedesign2}
\end{figure}

We check in Fig.~\ref{ch4:phase} that we recover the behaviour of Eq.~(\ref{eq:geom_phase}) (shown in dashed black line)
by simulating the phase-shift induced by a periodic grating of such nanoantennas all rotated by the same angle $\phi$ (green circles).
%and an efficiency in achieving the inversion of polarization of $90\%$ at normal incidence using \emph{Reticolo} \cite{hugonin2005reticolo},
%which can be explained by imperfect phase-profile due to phase discretization and losses by absorption.
%The difference between this efficiency $|R_{++}|^2=10\%$ and the criterion $|R_{+-}|^2=0\%$ can be explained by the imperfect phase-profile due to phase discretization.
We show in Fig.~\ref{ch4:metasurfacedesign2} the 3-D drawing of such a metasurface working at $\{\lambda_0=852\,\text{nm}, d=10\lambda_0\}$.
In this Figure, we also highlight the first super-cell (white box) starting from the center of the metasurface.

\begin{figure}[h!]
   \centering
   \includegraphics[width=\linewidth]{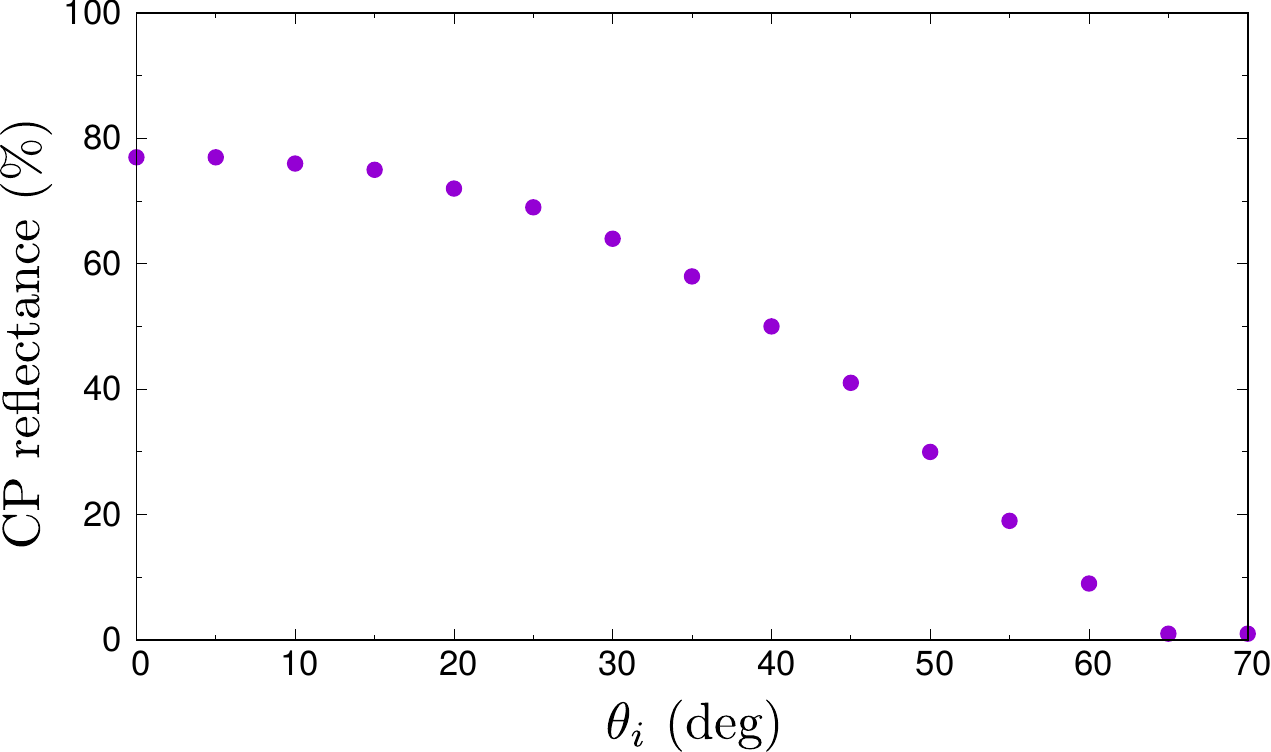}
   \caption{Cross-polarization (CP) power reflectance, which characterizes the conversion efficiency in energy between a light circularly polarized $\sigma^+$ and a light circularly polarized $\sigma^-$,
     as a function of the incident angle $\theta_i$, computed for a 2-D grating (see main text). The number of Fourier modes used to compute the CP reflectance is 30.}
   \label{ch4:metasurfacedesign2_eff}
\end{figure}

Next, in Fig.~\ref{ch4:metasurfacedesign2_eff}, we compute the conversion efficiency of a circularly polarized $\sigma^+$ incident wave reflected into a $\sigma^-$ circularly polarized wave  (cross-polarization reflectance) of the same 2-D grating as a function of the incident angle $\theta_i$. One can see that the cross-polarization power reflectance remains $>40\%$ for $\theta_i<45\degree$.
This design does not seem to be as good as the first design presented in Section~\ref{sec:resonant-phase-delay}, for which we recall that
the power reflectance into the desired order remains $>40\%$ up to $\theta_i=70\degree$ [Fig.~\ref{ch4:angle_supercell} (b)].
Even though the quantities that we compare here are different, both characterize in a way the performances of the metasurface.
%These quantities, that is the reflectance and the conversion efficiency between a light circularly polarized $\sigma^+$ and a light circularly polarized $\sigma^-$,
%are different (and one is computed for a grating of super-cells while the other one is computed for a grating of identical unit-cells) but they
%both characterize the performances of the metasurface, and it is what we eventually want to compare.
%All this is close to satisfying the criterion (ii).
%Here again, the difference between $|R_{+-}|^2=76\%$ and the criterion $|R_{+-}|^2=100\%$ can be explained by the phase discretization errors and also by the losses due to absorption (only, as we checked that the presence of the thick gold mirror prevents totally any transmission).

\section{Estimation of the coherence induced by the metasurface}
\label{ch4:sec:limitations}

In this Section, we want to assess a realistic value of the ground state coherence $\rho_{12}$ in the steady state [Eq.~(\ref{ch4:eq:coh_infty})] induced by a metasurface.
To do so, we limit the discussion to the first design (Section~\ref{sec:resonant-phase-delay}), since its performances seem to be better than for the second design (Section \ref{sec:geometa}).

If the dipole moments of the $\Lambda$-transition are equal ($d_{01}=d_{02}=d$), Eq.~(\ref{ch4:rhocartesian}) [or equivalently Eq.~(\ref{ch4:rhocylindrical})] becomes:
\begin{equation}
\rho_{12}(\infty)=\frac{1}{2}\times\frac{\text{Im}\left[G_{xx}-G_{yy}\right]}{\text{Im}\left[G_{xx}+G_{yy}\right]}\; ,
\label{ch4:rhocartesianrecall}
\end{equation}
where we recall that we consider $\text{Im}\left[G_{xy}\right]=0$ since the nanoantennas have a mirror symmetry.
By noting that, for a \emph{two-level atom} characterized by a dipole moment $\mathbf{d}$ oriented along the $x$-axis, the decay rate is given by \cite{lassalle2018interplay}:
\begin{equation}
  \gamma_x=\frac{2\omega_0^2}{\hbar\epsilon_0 c^2}\,|\mathbf{d}|^2\text{Im}\left(G_{xx}\right)\; ,
  \label{eq:decay_rate_green}
\end{equation}
and similarly for an orientation along the $y$-axis, Eq.~(\ref{ch4:rhocartesianrecall}) can be recast in the form:
\begin{equation}
\rho_{12}(\infty)=\frac{1}{2}\times\frac{\gamma_{x}-\gamma_{y}}{\gamma_{x}+\gamma_{y}}\; .
\label{ch4:rhocartesianrecallnew}
\end{equation}

To evaluate the coherence of Eq.~(\ref{ch4:rhocartesianrecallnew}), one could use Eq.~(\ref{eq:decay_rate_green})
  (or equivalently the expression in terms of the scattered field as in Refs.~\cite{jha2017metasurface,jha2018spontaneous})
  and numerically compute the Green tensor $\hat{\mathbf{G}}$ (resp. the scattered field $\mathbf{E}_s$) of the metasurface at the position of
  the atom, which is demanding in terms of computation time.
  %(the metasurface area is typically $10\lambda_0\times 10\lambda_0$).
  Instead, we choose to take advantage of the analytical results available for a spherical mirror \cite{hetet2010qed}. We consider that the quantity $\gamma_y$ is not modified compared to its free space value: $\gamma_y=\gamma_0$, while $\gamma_x$ is altered since the metasurface acts as a spherical mirror for such a polarization
  (in the case of the first design). The alteration of $\gamma_x$ is calculated as a function of the power reflectance $R_x$ (the subscript is for a light polarized along $\vec{x}$)
and the numerical aperture NA of the metasurface using the following expression \cite{hetet2010qed}:
\begin{equation}
\frac{\gamma_x}{\gamma_0}=3\int_{2\pi}\frac{\mathrm{d}\Omega}{4\pi}\left[1-\frac{|\mathbf{d}\cdot\mathbf{\Omega}|^2}{|\mathbf{d}|^2}\right]\times\left(1-R_x\right),
\label{sec:discuss:gamma1}
\end{equation}
where $\mathbf{\Omega}$ is the vectorial solid angle,
  and for the power reflectance $R_x$ in diffraction order $m=-1$, we take the values given in Table~\ref{ch4:tab:supercell}.
  $R_x$ is therefore a piecewise function, where its value, for a given location on the metasurface, is given by the underlying super-cell.
  Moreover, $R_x=0$ if $\text{sin}\theta>\text{NA}$, which takes into account the limited size of the metasurface.
  Here, the use of Eq.~(\ref{sec:discuss:gamma1}), which was originally derived for a two-level atom located at the focus of a spherical mirror in Ref.~\cite{hetet2010qed},
  is justified since the metasurface acts as a spherical mirror for a dipole emitter oriented along $\vec{x}$.
  In other words, the metasurface is \emph{optically equivalent} to a spherical mirror, and has the same Green tensor or scattered field value at the position of the atom;
  generally speaking, since the decay rate modification depends on these quantities (see Eq.~(\ref{eq:decay_rate_green}) in term of the Green tensor
  or Refs.~\cite{jha2017metasurface,jha2018spontaneous} in term of the scattered field),
  it should then be altered in the same way as in the case of a spherical mirror, in the far-field limit.
    At the present time, because of limitations of the computational resources, we were not able to provide a fully numerical estimate of the absolute coherence
    taking into account all the details of the metasurface. Actually, although we made use of Eq.~(\ref{sec:discuss:gamma1}) out of its original context (spherical mirror),
    we incorporated results from numerical computations (reflectance values in Table~\ref{ch4:tab:supercell}),
    which constitutes in our eyes an acceptable compromise, in-between a fully numerical treatment and an educated guess.

\begin{figure}[h!]
   \centering
   \includegraphics[width=\linewidth]{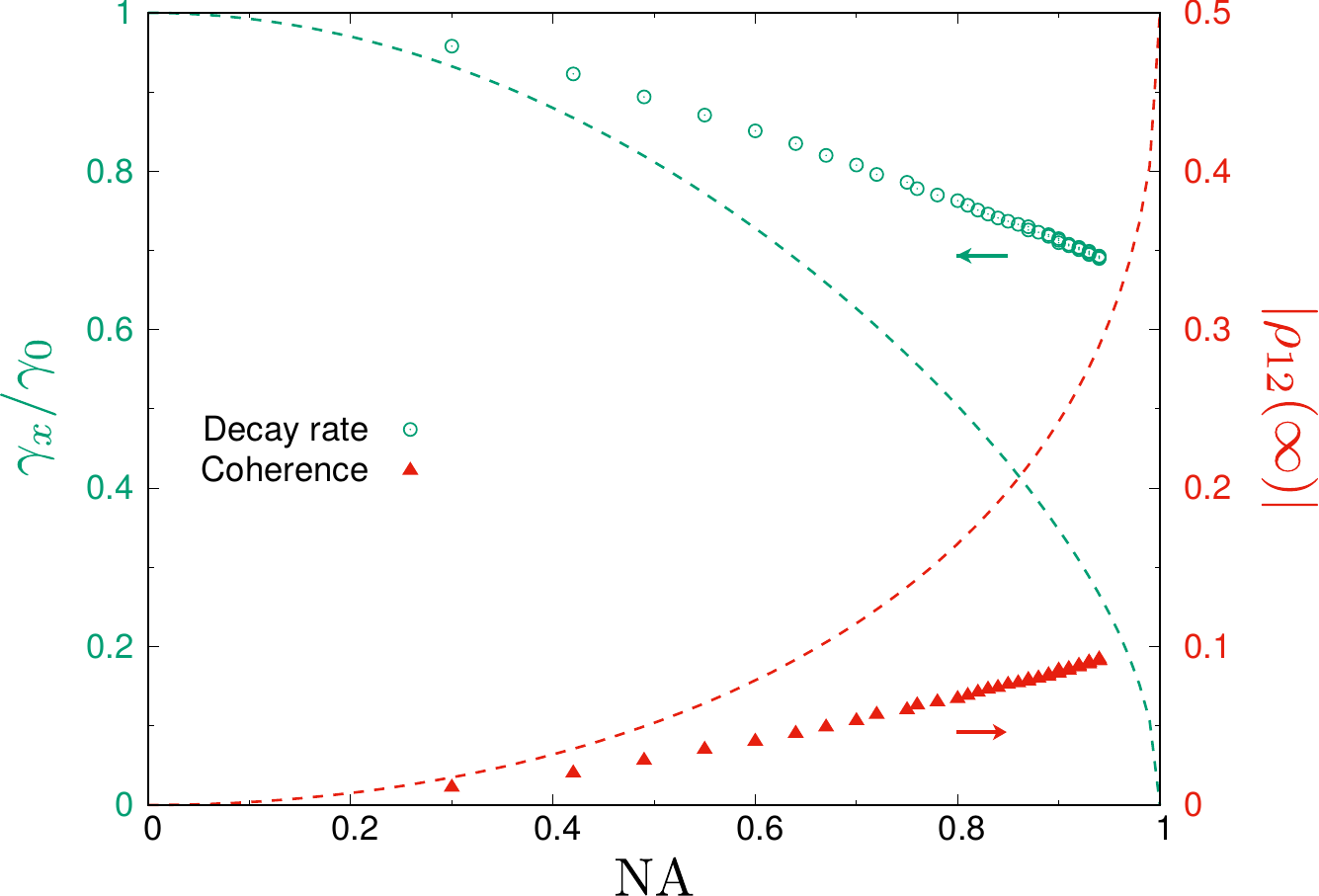}
   \caption{Relative decay rate modifications $\gamma_x/\gamma_0$ (green circles) and absolute coherence $|\rho_{12}|$ (red triangles) as a function of the numerical
     aperture of the metasurface NA. For comparison, the relative decay rate [resp. coherence] for an ideal spherical mirror of power reflectance $R_x=1$ for the $x$-polarization only is also shown (green [resp. red] dashed line).}
   \label{ch4:lastfig}
\end{figure}

In Fig.~\ref{ch4:lastfig}, we show the relative decay rate modifications $\gamma_x/\gamma_0$ calculated from Eq.~(\ref{sec:discuss:gamma1}) (green circles)
and the induced coherence calculated from Eq.~(\ref{ch4:rhocartesianrecallnew}) (red triangles), as a function of the numerical aperture defined as $\text{NA}\equiv\text{sin}\,\theta$. 
For comparison, we also show the decay rate modifications (resp. the induced coherence) in the case of a perfect reflective spherical mirror (power reflectance $R_x=1$) that would only reflect a polarization along $\vec{x}$ [green dashed line (resp. red dashed line)]. In this case, Eq.~(\ref{sec:discuss:gamma1}) can be calculated analytically and reads:
\begin{equation}
\frac{\gamma_x}{\gamma_0}=\sqrt{1-\text{NA}^2}\times\left(1-\frac{\text{NA}^2}{4}\right)\; .
\label{sec:discuss:gamma}
\end{equation}

One can see that for a metasurface of $\text{NA}= 0.7$, the decay rate $\gamma_x$ is reduced by 20\% compared to $\gamma_0$, with an induced coherence of $\sim 0.05$.
%\textit{i.e.} a reduction of about a factor of two with expected value of $37\%$ of an ideal lossless reflector {\color{red}(DW: check the value)}.
Compared to the ideal case of an infinite perfect spherical mirror, this value of the coherence is about one order of magnitude smaller ($0.5$ for an ideal reflector with NA$=1$).
Larger NA results only in a moderate improvement of the effect because of the rapid drop of the reflectance, contrary to the ideal case.
To attain near-unity efficiency in reflection, further optimizations of the antenna geometries that can take into consideration the coupling between neighboring elements are required.
Several methods have been proposed including objective-first algorithms \cite{ong2017freestanding,jafar2018adaptive,schmitt2019optimization}, topology optimization \cite{yang2017topology} and inverse designs \cite{piggott2014inverse,callewaert2018inverse},
which are also applicable to improve our device efficiency notably at large deflection angles, but beyond the scope of the present publication.

\section{Conclusion}

In this work, we predict the creation of a long-lifetime coherence between the two ground states of a quantum emitter with a $\Lambda$-configuration, induced by a quantum anisotropic vacuum (AQV).
An AQV can be engineered over macroscopic distances by a metasurface, made of metallic subwavelength reflectarrays and having a polarization-dependent response.
We proposed and designed two of such metasurfaces, based on the phase-mapping approach, using two different techniques: \emph{resonant phase-delays} and \emph{geometric phases}.
We quantify the efficiency of these metasurfaces to redirect the light on the quantum emitter, located at remote distances, by taking into account
the limitations on the numerical aperture due to the phase-mapping approach. Based on the exact results available for a spherical mirror,
we estimate a redirection of the light of about $20\%$ for a numerical aperture of $0.7$, leading to a coherence of $0.05$, which is one order of magnitude smaller than
in the ideal case of an infinite and perfect reflector.
Nevertheless, due to the long-lifetime of this coherence involving the ground states in a $\Lambda$-transition, this system allows for high resolution experiments,
and this effect should be observable using the current state-of-art NV-center experimental platform \cite{togan2010quantum}.
Detecting this coherence would represent an experimental demonstration of the anisotropy of vacuum on quantum emitters at remote distances.
In addition, this experiment would be a new test of quantum electrodynamics, in a counter-intuitive regime where coherence is driven by relaxation processes and vacuum fluctuations.
Moreover, such an experimental demonstration would also pave the way for controlling interactions between several quantum emitters
by the means of metasurfaces, which ultimately could be used to generate entanglement for quantum technology applications in a new paradigm \cite{hughes2017anisotropy,jha2017metasurface}.

\section*{Acknowledgements}
The authors wish to thank Gabriel H\'{e}tet, Guanghui Yuan, Giorgio Adamo, Weibo Gao and Martial Ducloy for fruitful discussions.
E. L. thanks Institut Fresnel (\emph{via} Fonds pour la Science 2018) and Nanyang Technological University for supporting his stay in Singapore.
This work is supported by the Singapore Ministry of Education Academic Research Fund Tier 3 Grant No. MOE2016-T3-1-006(S).

\setcounter{section}{0}
\renewcommand{\thesection}{\Alph{section}}
\renewcommand{\thesubsection}{\arabic{subsection}}
\setcounter{equation}{0}
\renewcommand{\theequation}{A\arabic{equation}}

\section*{Appendix: Master Equation derivation}

In this Appendix, we present the Master Equation framework, closely following Refs.~\cite{barnett2002methods} Chapter 5.6 and \cite{carmichael2013statistical} Chapter 1,
that we used to derive the master equation [Eq.~(\ref{chaqv:eq:master_eq_main})] in Section~\ref{ch4:section:theory}.

\subsection{Short notations}
It will be convenient for the following calculations to rewrite $\hat{H}_I(t)$ of Eq.~(\ref{chaqv:intH}) in a more compact form: 
\begin{equation}
\hat{H}_I(t) =
\hat{\mathbf{d}}^\dagger(t)\cdot\hat{\mathbf{E}}_v^{(+)}(t)
+ \hat{\mathbf{d}}(t)\cdot\hat{\mathbf{E}}_v^{(-)}(t)
\label{eq:H_short}
\end{equation}
where $\hat{\mathbf{d}}(t)$ and $\hat{\mathbf{d}}^\dagger(t)$ are
defined by:
\begin{equation}
\hat{\mathbf{d}}(t) \equiv - \left(\mathbf{d}_{01}^*\ket{1}\bra{0}e^{-\mathrm{i}\omega_1 t}+\mathbf{d}_{02}^*\ket{2}\bra{0}e^{-\mathrm{i}\omega_2 t}\right)
\label{eq:d}
\end{equation}
\begin{equation}
\hat{\mathbf{d}}^\dagger(t)\equiv -\left(\mathbf{d}_{01}\ket{0}\bra{1}e^{\mathrm{i}\omega_1 t}+\mathbf{d}_{02}\ket{0}\bra{2}e^{\mathrm{i}\omega_2 t}\right)
\label{eq:d_dagg}
\end{equation}
Note that for clarity we dropped the label $\bold{r}_0$
appearing in $\hat{\mathbf{E}}_v^{(+)}(\mathbf{r}_0,t)$ and
$\hat{\mathbf{E}}_v^{(-)}(\mathbf{r}_0,t)$, but one must remember that
the fields are evaluated at the position of the atom $\bold{r}_0$.
One must also take note that this Hamiltonian is written in the electric dipole and rotating-wave approximations.

\bigskip
\subsection{Master equation framework}
In the \emph{interaction picture}, the density matrix $\rho_T(t)$ of the total system
\{atom+environment\} obeys the Schr\"odinger equation \cite{barnett2002methods,carmichael2013statistical}:
\begin{equation}
\frac{\partial \rho_T(t)}{\partial t} =
\frac{1}{\mathrm{i}\hbar}[\hat{H}_I(t),\rho_T(t)]
\label{eq:rho_tot}
\end{equation}
The atomic density matrix $\rho(t)$ is obtained
by taking the trace over the degrees of freedom of the environment:
$\rho(t) = \text{Tr}_e(\rho_T(t))$, and therefore obeys:
\begin{equation}
\frac{\partial \rho(t)}{\partial t} =
\frac{1}{\mathrm{i}\hbar}\text{Tr}_e[\hat{H}_I(t),\rho_T(t)]
\label{eq:rho}
\end{equation}
We formally integrate Eq.~(\ref{eq:rho_tot}):
\begin{equation}
\rho_T(t) = \rho_T(0) +\frac{1}{\mathrm{i}\hbar}\int_0^t\mathrm{d}t'\,[\hat{H}_I(t'),\rho_T(t')]
\end{equation}
and substitute this expression in Eq.~(\ref{eq:rho}):
\begin{multline}
\frac{\partial \rho(t)}{\partial t} = \frac{1}{\mathrm{i}\hbar}\text{Tr}_e[\hat{H}_I(t),\rho_T(0)]\\-\frac{1}{\hbar^2}\int_0^t\mathrm{d}t'\,\text{Tr}_e\left[\hat{H}_I(t),[\hat{H}_I(t'),\rho_T(t')]\right]
\label{eq:rho_ME_before}
\end{multline}

Assuming that $\text{Tr}_e([\hat{H}_I(t),\rho_T(0)])=0$, we make the \emph{Born approximation}: $\rho_T(t)=\rho(t)\otimes\rho_e(0)$,
so that Eq.~(\ref{eq:rho_ME_before}) reduces to:
\begin{equation}
\frac{\partial \rho(t)}{\partial t} = -\frac{1}{\hbar^2}\int_0^t\mathrm{d}t'\,\text{Tr}_e\left[\hat{H}_I(t),[\hat{H}_I(t'),\rho_e(0)\otimes\rho(t')]\right]
\label{eq:rho_ME_bis}
\end{equation}
Next, we make the \emph{Markov approximation} and replace $\rho(t')$ by
$\rho(t)$ in the integrand. 
Therefore, we get a Master Equation for the atomic density matrix $\rho(t)$ in the \emph{Born-Markov approximation}:
\begin{equation}
\frac{\partial \rho(t)}{\partial t} = -\frac{1}{\hbar^2}\int_0^t\mathrm{d}t'\,\text{Tr}_e\left[\hat{H}_I(t),[\hat{H}_I(t'),\rho_e(0)\otimes\rho(t)]\right]
\label{eq:rho_ME}
\end{equation}

Now, we write $\hat{H}_I(t)$
explicitely and expand the commutators. Using the compact form Eq.~(\ref{eq:H_short}) into Eq.~(\ref{eq:rho_ME}) one gets:
\begin{widetext}
\begin{equation}
\frac{\partial \rho(t)}{\partial t} = -\frac{1}{\hbar^2}\int_0^t\mathrm{d}t'\,\text{Tr}_e\left[\hat{\mathbf{d}}^\dagger(t)\cdot\hat{\mathbf{E}}_v^{(+)}(t)
+ \hat{\mathbf{d}}(t)\cdot\hat{\mathbf{E}}_v^{(-)}(t),[\hat{\mathbf{d}}^\dagger(t')\cdot\hat{\mathbf{E}}_v^{(+)}(t')
+
\hat{\mathbf{d}}(t')\cdot\hat{\mathbf{E}}_v^{(-)}(t'),\rho_e(0)\otimes\rho(t)]\right]
\label{eq_ME_calc}
\end{equation}
\end{widetext}
Expanding the commutators in
Eq.~(\ref{eq_ME_calc}) gives 16 terms. Noting that the trace only acts on the field
operators and on $\rho_e(0)$, and using the cyclic property of the
trace operation and the fact that for instance $\text{Tr}_e\left(\rho_e(0))\hat{\mathbf{E}}_v^{(+)}(t)\hat{\mathbf{E}}_v^{(-)}(t')\right)
= \left<\hat{\mathbf{E}}_v^{(+)}(t)\hat{\mathbf{E}}_v^{(-)}(t')\right>$, we find

\begin{multline}
\frac{\partial \rho(t)}{\partial t} =
  -\frac{1}{\hbar^2}\int_0^t\mathrm{d}t'\,\\
  \left<\hat{\mathbf{E}}_v^{(+)}(t)\hat{\mathbf{E}}_v^{(-)}(t')\right>\left(
    \hat{\mathbf{d}}^\dagger(t)\hat{\mathbf{d}}(t')\rho(t)-\hat{\mathbf{d}}(t')\rho(t)\hat{\mathbf{d}}^\dagger(t)\right)\\
+ \left<\hat{\mathbf{E}}_v^{(+)}(t')\hat{\mathbf{E}}_v^{(-)}(t)\right>\left(
    \rho(t)\hat{\mathbf{d}}^\dagger(t')\hat{\mathbf{d}}(t)-\hat{\mathbf{d}}(t)\rho(t)\hat{\mathbf{d}}^\dagger(t')\right)\\
+ \left<\hat{\mathbf{E}}_v^{(-)}(t)\hat{\mathbf{E}}_v^{(+)}(t')\right>\left(
    \hat{\mathbf{d}}(t)\hat{\mathbf{d}}^\dagger(t')\rho(t)-\hat{\mathbf{d}}^\dagger(t')\rho(t)\hat{\mathbf{d}}(t)\right)\\
+ \left<\hat{\mathbf{E}}_v^{(-)}(t')\hat{\mathbf{E}}_v^{(+)}(t)\right>\left(
    \rho(t)\hat{\mathbf{d}}(t')\hat{\mathbf{d}}^\dagger(t)-\hat{\mathbf{d}}^\dagger(t)\rho(t)\hat{\mathbf{d}}(t')\right)\\
+ \left<\hat{\mathbf{E}}_v^{(+)}(t)\hat{\mathbf{E}}_v^{(+)}(t')\right>\left(
    \hat{\mathbf{d}}^\dagger(t)\hat{\mathbf{d}}^\dagger(t')\rho(t)-\hat{\mathbf{d}}^\dagger(t')\rho(t)\hat{\mathbf{d}}^\dagger(t)\right)\\
+ \left<\hat{\mathbf{E}}_v^{(+)}(t')\hat{\mathbf{E}}_v^{(+)}(t)\right>\left(
    \rho(t)\hat{\mathbf{d}}^\dagger(t')\hat{\mathbf{d}}^\dagger(t)-\hat{\mathbf{d}}^\dagger(t)\rho(t)\hat{\mathbf{d}}^\dagger(t')\right)\\
+ \left<\hat{\mathbf{E}}_v^{(-)}(t)\hat{\mathbf{E}}_v^{(-)}(t')\right>\left(
    \hat{\mathbf{d}}(t)\hat{\mathbf{d}}(t')\rho(t)-\hat{\mathbf{d}}(t')\rho(t)\hat{\mathbf{d}}(t)\right)\\
+ \left<\hat{\mathbf{E}}_v^{(-)}(t')\hat{\mathbf{E}}_v^{(-)}(t)\right>\left(
    \rho(t)\hat{\mathbf{d}}(t')\hat{\mathbf{d}}(t)-\hat{\mathbf{d}}(t)\rho(t)\hat{\mathbf{d}}(t')\right)\\
\label{eq:rho_ME_develop}
\end{multline}

We make the two following additional approximations:
\begin{itemize}
\item
$\left<\hat{\mathbf{E}}_v^{(+)}(t)\hat{\mathbf{E}}_v^{(+)}(t')\right>
= \left<\hat{\mathbf{E}}_v^{(+)}(t')\hat{\mathbf{E}}_v^{(+)}(t)\right>
= \left<\hat{\mathbf{E}}_v^{(-)}(t)\hat{\mathbf{E}}_v^{(-)}(t')\right>
= \left<\hat{\mathbf{E}}_v^{(-)}(t')\hat{\mathbf{E}}_v^{(-)}(t)\right>
= 0$ \\(which is valid for an environment in thermodynamic equilibrium)
\item
  $\left<\hat{\mathbf{E}}_v^{(-)}(t)\hat{\mathbf{E}}_v^{(+)}(t')\right>=\left<\hat{\mathbf{E}}_v^{(-)}(t')\hat{\mathbf{E}}_v^{(+)}(t)\right>=0$
  \\(which is valid for optical frequencies)
\end{itemize}

Thus, only the first two terms remain in Eq.~(\ref{eq:rho_ME_develop})
which reduces to
\begin{multline}
\frac{\partial \rho(t)}{\partial t} =
  -\frac{1}{\hbar^2}\int_0^t\mathrm{d}t'\,\\
  \left<\hat{\mathbf{E}}_v^{(+)}(t)\hat{\mathbf{E}}_v^{(-)}(t')\right>\left(
    \hat{\mathbf{d}}^\dagger(t)\hat{\mathbf{d}}(t')\rho(t)-\hat{\mathbf{d}}(t')\rho(t)\hat{\mathbf{d}}^\dagger(t)\right)\\
+ \left<\hat{\mathbf{E}}_v^{(+)}(t')\hat{\mathbf{E}}_v^{(-)}(t)\right>\left(
    \rho(t)\hat{\mathbf{d}}^\dagger(t')\hat{\mathbf{d}}(t)-\hat{\mathbf{d}}(t)\rho(t)\hat{\mathbf{d}}^\dagger(t')\right)\\
\label{eq:rho_ME_simp}
\end{multline}

By using the property of the correlation function $\left<\hat{\mathbf{E}}_v^{(+)}(t')\hat{\mathbf{E}}_v^{(-)}(t)\right>=\left<\hat{\mathbf{E}}_v^{(+)}(t)\hat{\mathbf{E}}_v^{(-)}(t')\right>^*$,
\\and noting the fact that $\left(
    \rho(t)\hat{\mathbf{d}}^\dagger(t')\hat{\mathbf{d}}(t)-\hat{\mathbf{d}}(t)\rho(t)\hat{\mathbf{d}}^\dagger(t')\right)
  = \left(
    \hat{\mathbf{d}}^\dagger(t)\hat{\mathbf{d}}(t')\rho(t)-\hat{\mathbf{d}}(t')\rho(t)\hat{\mathbf{d}}^\dagger(t)\right)^\dagger$,\\
one can see that the second term in Eq.~(\ref{eq:rho_ME_simp}) is
actually the Hermitian conjugate (H.c.) of the first one. Therefore, we
simply write Eq.~(\ref{eq:rho_ME_simp}) as
\begin{widetext}
\begin{equation}
\frac{\partial \rho(t)}{\partial t} =
  -\frac{1}{\hbar^2}\int_0^t\mathrm{d}t'\,\left<\hat{\mathbf{E}}_v^{(+)}(t)\hat{\mathbf{E}}_v^{(-)}(t')\right>\left(
    \hat{\mathbf{d}}^\dagger(t)\hat{\mathbf{d}}(t')\rho(t)-\hat{\mathbf{d}}(t')\rho(t)\hat{\mathbf{d}}^\dagger(t)\right)
+ \text{H.c.}
\label{eq:rho_ME_Hc2}
\end{equation}
\end{widetext}

\subsection{Master equation for an atomic $\Lambda$-transition}

Eq.~(\ref{eq:rho_ME_Hc2}) is the starting point to calculate the dynamical evolution of any multilevel atom.
Here, we proceed by writing explicitely the terms in the integrand using the
expressions for $\hat{\mathbf{d}}(t)$ and
$\hat{\mathbf{d}}^\dagger(t)$ from Eqs.~(\ref{eq:d}) and (\ref{eq:d_dagg}), which corresponds to the $\Lambda$-configuration with orthogonal transitions:
\begin{multline}
\hat{\mathbf{d}}^\dagger(t)\hat{\mathbf{d}}(t')\rho(t)=
  e^{\mathrm{i}\omega_1(t-t')}\bold{d}_{01}\ket{0}\bra{0}\bold{d}_{01}^* \rho(t)\\
+e^{\mathrm{i}\omega_2(t-t')}\bold{d}_{02}\ket{0}\bra{0}\bold{d}_{02}^* \rho(t)
\end{multline}
and by defining $\rho_{00}(t)\equiv \bra{0}\rho(t)\ket{0}$ to simplify the expressions:
\begin{multline}
\hat{\mathbf{d}}(t')\rho(t)\hat{\mathbf{d}}^\dagger(t)=
+e^{\mathrm{i}\omega_1(t-t')}\rho_{00}(t)\bold{d}_{01}^*\ket{1}\bra{1}\bold{d}_{01}\\
+e^{\mathrm{i}\omega_2(t-t')}\rho_{00}(t)\bold{d}_{02}^*\ket{2}\bra{2}\bold{d}_{02}\\
+e^{\mathrm{i}\omega_1t}e^{-\mathrm{i}\omega_2t'}\rho_{00}(t)\bold{d}_{02}^*\ket{2}\bra{1}\bold{d}_{01}\\
+e^{\mathrm{i}\omega_2t}e^{-\mathrm{i}\omega_1t'}\rho_{00}(t)\bold{d}_{01}^*\ket{1}\bra{2}\bold{d}_{02}\\
\end{multline}

Substituting these expressions in Eq.~(\ref{eq:rho_ME_Hc2}) and
by factorizing the exponential terms, we get:
\begin{multline}
\frac{\partial \rho(t)}{\partial t} =
  -\frac{1}{\hbar^2}\int_0^t\mathrm{d}t'\,\left<\hat{\mathbf{E}}_v^{(+)}(t)\hat{\mathbf{E}}_v^{(-)}(t')\right>\times\\
+e^{\mathrm{i}\omega_1(t-t')}\left(\bold{d}_{01}\ket{0}\bra{0}\bold{d}_{01}^* \rho(t)
  - \rho_{00}(t)\bold{d}_{01}^*\ket{1}\bra{1}\bold{d}_{01}\right)\\
+e^{\mathrm{i}\omega_2(t-t')}\left(
  \bold{d}_{02}\ket{0}\bra{0}\bold{d}_{02}^* \rho(t)-\rho_{00}(t)\bold{d}_{02}^*\ket{2}\bra{2}\bold{d}_{02}\right)\\
-e^{\mathrm{i}\omega_1t}e^{-\mathrm{i}\omega_2t'}\rho_{00}(t)\bold{d}_{02}^*\ket{2}\bra{1}\bold{d}_{01}\\
-
e^{\mathrm{i}\omega_2t}e^{-\mathrm{i}\omega_1t'}\rho_{00}(t)\bold{d}_{01}^*\ket{1}\bra{2}\bold{d}_{02}\\
+\text{H.c.}\\
\label{eq:rho_ME_Hc3}
\end{multline}

We now write: $\left<\hat{\mathbf{E}}_v^{(+)}(t)\hat{\mathbf{E}}_v^{(-)}(t')\right>=\left<\hat{\mathbf{E}}_v^{(+)}(t-t')\hat{\mathbf{E}}_v^{(-)}(0)\right>$
(the correlation function only depends on the time difference). Making the change of variable $\tau=t-t'$, the next approximation is to make the upper
limit tend to infinity. Eq.~(\ref{eq:rho_ME_Hc3}) becomes
\begin{multline}
\frac{\partial \rho(t)}{\partial t} =
  -\frac{1}{\hbar^2}\int_0^\infty\mathrm{d}\tau\,\left<\hat{\mathbf{E}}_v^{(+)}(\tau)\hat{\mathbf{E}}_v^{(-)}(0)\right>\times\\
+e^{\mathrm{i}\omega_1\tau}\left(\bold{d}_{01}\ket{0}\bra{0}\bold{d}_{01}^* \rho(t)
  - \rho_{00}(t)\bold{d}_{01}^*\ket{1}\bra{1}\bold{d}_{01}\right)\\
+e^{\mathrm{i}\omega_2\tau}\left(
  \bold{d}_{02}\ket{0}\bra{0}\bold{d}_{02}^* \rho(t)-\rho_{00}(t)\bold{d}_{02}^*\ket{2}\bra{2}\bold{d}_{02}\right)\\
-e^{\mathrm{i}(\omega_1-\omega_2)t}e^{\mathrm{i}\omega_2\tau}\rho_{00}(t)\bold{d}_{02}^*\ket{2}\bra{1}\bold{d}_{01}\\
-
e^{\mathrm{i}(\omega_2-\omega_1)t}e^{\mathrm{i}\omega_1\tau}\rho_{00}(t)\bold{d}_{01}^*\ket{1}\bra{2}\bold{d}_{02}\\
+\text{H.c.}\\
\label{eq:rho_ME_Hc4}
\end{multline}

We finally introduce the positive part of the correlation tensor as:
\begin{equation}
\hat{\mathbf{C}}^{(+)}(\omega)\equiv\int_0^\infty\mathrm{d}\tau\,\left<\hat{\mathbf{E}}_v^{(+)}(\tau)\hat{\mathbf{E}}_v^{(-)}(0)\right>e^{\mathrm{i}\omega\tau}
\end{equation}
to get:
\begin{multline}
\frac{\partial \rho(t)}{\partial t} = -\Gamma_1\left(
  \ket{0}\bra{0}\rho(t) -\rho_{00}(t)\ket{1}\bra{1}\right) \\-\Gamma_2\left(
  \ket{0}\bra{0}\rho(t)-\rho_{00}(t)\ket{2}\bra{2}\right)\\
+\Gamma_{21}e^{\mathrm{i}(\omega_1-\omega_2)t}\rho_{00}(t)\ket{2}\bra{1}\\
+\Gamma_{12}e^{\mathrm{i}(\omega_2-\omega_1)t}\rho_{00}(t)\ket{1}\bra{2}\\
+\text{H.c.}\\
\end{multline}
with the following definitions of the coefficients:
\begin{equation}
\Gamma_i \equiv \frac{1}{\hbar^2}\bold{d}_{0i}^*\cdot
\hat{\mathbf{C}}^{(+)}(\omega_i)\cdot\bold{d}_{0i}
\end{equation}
and 
\begin{equation}
\Gamma_{ij} \equiv \frac{1}{\hbar^2}\bold{d}_{0i}^*\cdot
\hat{\mathbf{C}}^{(+)}(\omega_i)\cdot\bold{d}_{0j}
\label{eq:kappa}
\end{equation}

Remember that in the Master Equation above $\rho(t)$ is still in the interaction picture, and
we come back to the \emph{Schr\"odinger picture} assuming furthermore that the transition energies are about the same $\omega_1\simeq\omega_2\equiv\omega_0$
\begin{multline}
\frac{\partial \rho(t)}{\partial t} = -\mathrm{i}\omega_0\ket{0}\bra{0}\rho(t) \\-\Gamma_1\left(
  \ket{0}\bra{0}\rho(t) -\rho_{00}(t)\ket{1}\bra{1}\right) \\-\Gamma_2\left(
  \ket{0}\bra{0}\rho(t) -\rho_{00}(t)\ket{2}\bra{2}\right)\\
+\Gamma_{21}\rho_{00}(t)\ket{2}\bra{1}
+\Gamma_{12}\rho_{00}(t)\ket{1}\bra{2}\\
+\text{H.c.}\\
\label{app:eq:master_eq_main}
\end{multline}

In Eq.~(\ref{app:eq:master_eq_main}), we have introduced the definitions of the coefficients:
\begin{equation}
\Gamma_i \equiv \frac{1}{\hbar^2}\bold{d}_{0i}^*\cdot
\hat{\mathbf{C}}^{(+)}(\mathbf{r}_0,\mathbf{r}_0,\omega_0)\cdot\bold{d}_{0i}
\end{equation}
and 
\begin{equation}
\Gamma_{ij} \equiv \frac{1}{\hbar^2}\bold{d}_{0i}^*\cdot
\hat{\mathbf{C}}^{(+)}(\mathbf{r}_0,\mathbf{r}_0,\omega_0)\cdot\bold{d}_{0j} 
\label{eq:kappa_main}
\end{equation}
defined in terms of the \emph{positive part of the correlation tensor} $\hat{\mathbf{C}}^{(+)}$ that reads:
\begin{equation}
\hat{\mathbf{C}}^{(+)}(\mathbf{r},\mathbf{r}',\omega)=\int_{0}^{+\infty}\mathrm{d}\tau\,\left<\hat{\mathbf{E}}_v^{(+)}(\mathbf{r},\tau)\hat{\mathbf{E}}_v^{(-)}(\mathbf{r}',0)\right>e^{\mathrm{i}\omega\tau}
\label{chaqv:eq:correltenspos}
\end{equation}
where the bracket indicates an ensemble average:
\begin{equation}
  \left<\hat{\mathbf{E}}_v^{(+)}(\mathbf{r},\tau)\hat{\mathbf{E}}_v^{(-)}(\mathbf{r}',0)\right>
  \equiv\text{Tr}_e\left(\rho_e(0)\hat{\mathbf{E}}_v^{(+)}(\mathbf{r},\tau)\hat{\mathbf{E}}_v^{(-)}(\mathbf{r}',0)\right)
\end{equation}
Using the mathematical relation:
\begin{equation}
\mathcal{P}\left(\frac{1}{x}\right)=\frac{1}{x+\mathrm{i}\epsilon}+\mathrm{i}\pi\delta(x) \quad\quad\text{with}\quad \epsilon\rightarrow 0
\end{equation}
one can demonstrate that:
\begin{equation}
  \hat{\mathbf{C}}^{(+)}(\mathbf{r},\mathbf{r}',\omega_0)=\frac{1}{2}\hat{\mathbf{C}}(\mathbf{r},\mathbf{r}',\omega_0)+
  \frac{\mathrm{i}}{2\pi}\mathcal{P}\left\{\int_{0}^{+\infty}\mathrm{d}\omega\,\frac{\hat{\mathbf{C}}(\mathbf{r},\mathbf{r}',\omega)}{\omega_0-\omega}\right\}
\end{equation}
where $\hat{\mathbf{C}}$ is the correlation tensor defined as:
\begin{equation}
\hat{\mathbf{C}}(\mathbf{r},\mathbf{r}',\omega)\equiv\int_{-\infty}^{+\infty}\mathrm{d}\tau\,\left<\hat{\mathbf{E}}_v^{(+)}(\mathbf{r},\tau)\hat{\mathbf{E}}_v^{(-)}(\mathbf{r}',0)\right>e^{\mathrm{i}\omega\tau}
\label{chaqv:eq:correltensposss}
\end{equation}
Therefore, the coefficients $\Gamma_i$ become:
\begin{equation}
\Gamma_i=\frac{\gamma_i}{2}+\mathrm{i}\Delta\omega_i
\end{equation}
with
\begin{equation}
\gamma_i= \frac{1}{\hbar^2}\bold{d}_{0i}^*\cdot
\hat{\mathbf{C}}(\mathbf{r}_0,\mathbf{r}_0,\omega_0)\cdot\bold{d}_{0i}
\end{equation}
and
\begin{equation}
\Delta\omega_i=\frac{1}{2\pi \hbar^2}\mathcal{P}\left\{\int_{0}^{+\infty}\mathrm{d}\omega\,\frac{\bold{d}_{0i}^*\cdot\hat{\mathbf{C}}(\mathbf{r}_0,\mathbf{r}_0,\omega)\cdot\bold{d}_{0i}}{\omega_0-\omega}\right\}
\end{equation}
where $\gamma_i$ can be interpreted as the decay rate on the transition $\ket{0}\rightarrow\ket{i}$, and $\Delta\omega_i$
is the Lamb shift of the level $\ket{i}$.

In the following, we recast the Lamb shift into the transition frequency and reduce $\hat{\mathbf{C}}^{(+)}(\mathbf{r},\mathbf{r}',\omega_0)$ as:
\begin{equation}
\hat{\mathbf{C}}^{(+)}(\mathbf{r},\mathbf{r}',\omega_0)\equiv\frac{1}{2}\hat{\mathbf{C}}(\mathbf{r},\mathbf{r}',\omega_0)
\end{equation}
Therefore, the coefficients become:
\begin{equation}
\Gamma_i=\frac{\gamma_i}{2}\quad\quad\text{with}\quad \gamma_i=\frac{1}{\hbar^2}\bold{d}_{0i}^*\cdot
\hat{\mathbf{C}}(\mathbf{r}_0,\mathbf{r}_0,\omega_0)\cdot\bold{d}_{0i}
\end{equation}
and
\begin{equation}
\Gamma_{ij}=\frac{\kappa_{ij}}{2}\quad\quad\text{with}\quad \kappa_{ij}=\frac{1}{\hbar^2}\bold{d}_{0i}^*\cdot
\hat{\mathbf{C}}(\mathbf{r}_0,\mathbf{r}_0,\omega_0)\cdot\bold{d}_{0j}.
\end{equation}

\bigskip
\subsection{Solution of the master equation}

From the Master Equation, given in Eq.~(\ref{app:eq:master_eq_main}), we obtain the following equations for the atomic populations $\rho_{ii}(t)$ and atomic coherences $\rho_{ij}(t)$ with $j\neq i$
\begin{equation}
\dot{\rho}_{ii}(t)=\gamma_i\rho_{00}(t)\quad\text{for}\quad i=1,2
\label{app:eq:rho_11}
\end{equation}
\begin{equation}
\dot{\rho}_{00}(t)=-(\gamma_1+\gamma_2)\rho_{00}(t)
\label{app:eq:rho_33}
\end{equation}
\begin{equation}
\dot{\rho}_{i0}(t)=-\left(\frac{\gamma_1+\gamma_2}{2}-\mathrm{i}\omega_0\right)\rho_{i0}(t)\quad\text{for}\quad i=1,2
\label{app:eq:rho_13}
\end{equation}
\begin{equation}
\dot{\rho}_{12}(t)=\kappa_{12}\rho_{00}(t)
\label{app:eq:coherences}
\end{equation}
where we used the fact that $\kappa_{21}^*=\kappa_{12}$.
Note that these equations are also supplemented by their conjugates.

The atom is initially prepared in the excited state with the following initial conditions (at $t=0$):
$\rho_{00}(0)=1$, $\rho_{11}(0)=\rho_{22}(0)= 0$ and $\rho_{ij}(0)=0$ for $j\neq i$.
%$\left\{
%        \begin{array}{ll} 
%      \rho_{11}(0)=\rho_{22}(0)= 0\\
%      \rho_{00}(0)=1\\
%      \rho_{ij}(0)=0 \,\,\text{for}\,\, j\neq i\\
%      \end{array}
%\right.$
Solving Eqs.~(\ref{app:eq:rho_11}) and (\ref{app:eq:rho_33}) with the initial
conditions above is straightforward. With the initial condition $\rho_{00}(0) = 1$, Eq.~(\ref{app:eq:rho_33})
gives
\begin{equation}
\rho_{00}(t)=e^{-(\gamma_1+\gamma_2)t}
\Rightarrow\rho_{00}(\infty) = 0
\label{app:eq:rho_33_bis}
\end{equation}
Substituting it in Eqs.~(\ref{app:eq:rho_11}) and carrying out the integration with the initial
conditions $\rho_{11}(0)=\rho_{22}(0)= 0$ gives
\begin{equation}
\rho_{11}(t)=\frac{\gamma_1}{\gamma_1+\gamma_2}\left[
  1-e^{-(\gamma_1+\gamma_2)t}\right] \Rightarrow \rho_{11}(\infty) = \frac{\gamma_1}{\gamma_1+\gamma_2}
\label{app:ch4:eq:pop1}
\end{equation}
\begin{equation}
\rho_{22}(t)=\frac{\gamma_2}{\gamma_1+\gamma_2}\left[
  1-e^{-(\gamma_1+\gamma_2)t}\right] \Rightarrow \rho_{22}(\infty) = \frac{\gamma_2}{\gamma_1+\gamma_2}
\label{app:ch4:eq:pop2}
\end{equation}
Furthermore, integration of Eq.~(\ref{app:eq:rho_13}) together
with the initial condition $\rho_{ij}(0)=0$ for $j\neq i$ gives
\begin{equation}
\rho_{10}(t)=\rho_{20}(t)=0 \,\,\,\,\, \forall t
\end{equation}
Finally, for the coherence $\rho_{12}(t)$ given by
Eq.~(\ref{app:eq:coherences}), substituting the expression of
$\rho_{00}(t)$ [Eq.~(\ref{app:eq:rho_33_bis})] in Eq.~(\ref{app:eq:coherences}) gives:
\begin{equation}
\dot{\rho}_{12}(t)=\kappa_{12}e^{-(\gamma_1+\gamma_2)t}
\end{equation}
and after integration, together with the initial condition
$\rho_{12}(0)=0$, we find 
\begin{equation}
  \rho_{12}(t)=\frac{\kappa_{12}}{\gamma_1+\gamma_2}\left[1-e^{-(\gamma_1+\gamma_2)t}\right]
  \label{app:ch4:eq:coh_time}
\end{equation}
and for $t\rightarrow \infty$
\begin{equation}
\rho_{12}(\infty)=\frac{\kappa_{12}}{\gamma_1+\gamma_2}
  \label{app:ch4:eq:coh_infty}
\end{equation}

%\section{Dressed-state approach}
%\label{app:dressed-state}
%\input{./Appendix/dressed-state.tex}

%\section{Metasurface design: super-cells}
%\label{ch5:app:sc}
%\input{./Appendix/super-cells.tex}

%\section{Metasurface design: unit-cells}
%\label{ch5:app:uc}
%\input{./Appendix/unit-cells.tex}

%\section{Design B: microscopic considerations}
%\label{app:designB}
%\input{./Appendix/designB.tex}

% Bibliography
\bibliographystyle{apsrev4-1} 
\bibliography{biblio}

\begin{thebibliography}{52}%
\makeatletter
\providecommand \@ifxundefined [1]{%
 \@ifx{#1\undefined}
}%
\providecommand \@ifnum [1]{%
 \ifnum #1\expandafter \@firstoftwo
 \else \expandafter \@secondoftwo
 \fi
}%
\providecommand \@ifx [1]{%
 \ifx #1\expandafter \@firstoftwo
 \else \expandafter \@secondoftwo
 \fi
}%
\providecommand \natexlab [1]{#1}%
\providecommand \enquote  [1]{``#1''}%
\providecommand \bibnamefont  [1]{#1}%
\providecommand \bibfnamefont [1]{#1}%
\providecommand \citenamefont [1]{#1}%
\providecommand \href@noop [0]{\@secondoftwo}%
\providecommand \href [0]{\begingroup \@sanitize@url \@href}%
\providecommand \@href[1]{\@@startlink{#1}\@@href}%
\providecommand \@@href[1]{\endgroup#1\@@endlink}%
\providecommand \@sanitize@url [0]{\catcode `\\12\catcode `\$12\catcode
  `\&12\catcode `\#12\catcode `\^12\catcode `\_12\catcode `\%12\relax}%
\providecommand \@@startlink[1]{}%
\providecommand \@@endlink[0]{}%
\providecommand \url  [0]{\begingroup\@sanitize@url \@url }%
\providecommand \@url [1]{\endgroup\@href {#1}{\urlprefix }}%
\providecommand \urlprefix  [0]{URL }%
\providecommand \Eprint [0]{\href }%
\providecommand \doibase [0]{http://dx.doi.org/}%
\providecommand \selectlanguage [0]{\@gobble}%
\providecommand \bibinfo  [0]{\@secondoftwo}%
\providecommand \bibfield  [0]{\@secondoftwo}%
\providecommand \translation [1]{[#1]}%
\providecommand \BibitemOpen [0]{}%
\providecommand \bibitemStop [0]{}%
\providecommand \bibitemNoStop [0]{.\EOS\space}%
\providecommand \EOS [0]{\spacefactor3000\relax}%
\providecommand \BibitemShut  [1]{\csname bibitem#1\endcsname}%
\let\auto@bib@innerbib\@empty
%</preamble>
\bibitem [{\citenamefont {Raimond}\ and\ \citenamefont
  {Haroche}(2006)}]{raimond2006exploring}%
  \BibitemOpen
  \bibfield  {author} {\bibinfo {author} {\bibfnamefont {J.-M.}\ \bibnamefont
  {Raimond}}\ and\ \bibinfo {author} {\bibfnamefont {S.}~\bibnamefont
  {Haroche}},\ }\href@noop {} {\emph {\bibinfo {title} {Exploring the
  quantum}}}\ (\bibinfo  {publisher} {Oxford University Press, Oxford},\
  \bibinfo {year} {2006})\BibitemShut {NoStop}%
\bibitem [{\citenamefont {Baranov}\ \emph {et~al.}(2017)\citenamefont
  {Baranov}, \citenamefont {Wersall}, \citenamefont {Cuadra}, \citenamefont
  {Antosiewicz},\ and\ \citenamefont {Shegai}}]{baranov2017novel}%
  \BibitemOpen
  \bibfield  {author} {\bibinfo {author} {\bibfnamefont {D.~G.}\ \bibnamefont
  {Baranov}}, \bibinfo {author} {\bibfnamefont {M.}~\bibnamefont {Wersall}},
  \bibinfo {author} {\bibfnamefont {J.}~\bibnamefont {Cuadra}}, \bibinfo
  {author} {\bibfnamefont {T.~J.}\ \bibnamefont {Antosiewicz}}, \ and\ \bibinfo
  {author} {\bibfnamefont {T.}~\bibnamefont {Shegai}},\ }\href@noop {}
  {\bibfield  {journal} {\bibinfo  {journal} {ACS Photonics}\ }\textbf
  {\bibinfo {volume} {5}},\ \bibinfo {pages} {24} (\bibinfo {year}
  {2017})}\BibitemShut {NoStop}%
\bibitem [{\citenamefont {H{\'e}tet}\ \emph {et~al.}(2010)\citenamefont
  {H{\'e}tet}, \citenamefont {Slodi{\v{c}}ka}, \citenamefont {Gl{\"a}tzle},
  \citenamefont {Hennrich},\ and\ \citenamefont {Blatt}}]{hetet2010qed}%
  \BibitemOpen
  \bibfield  {author} {\bibinfo {author} {\bibfnamefont {G.}~\bibnamefont
  {H{\'e}tet}}, \bibinfo {author} {\bibfnamefont {L.}~\bibnamefont
  {Slodi{\v{c}}ka}}, \bibinfo {author} {\bibfnamefont {A.}~\bibnamefont
  {Gl{\"a}tzle}}, \bibinfo {author} {\bibfnamefont {M.}~\bibnamefont
  {Hennrich}}, \ and\ \bibinfo {author} {\bibfnamefont {R.}~\bibnamefont
  {Blatt}},\ }\href@noop {} {\bibfield  {journal} {\bibinfo  {journal}
  {Physical Review A}\ }\textbf {\bibinfo {volume} {82}},\ \bibinfo {pages}
  {063812} (\bibinfo {year} {2010})}\BibitemShut {NoStop}%
\bibitem [{\citenamefont {Dorner}\ and\ \citenamefont
  {Zoller}(2002)}]{dorner2002laser}%
  \BibitemOpen
  \bibfield  {author} {\bibinfo {author} {\bibfnamefont {U.}~\bibnamefont
  {Dorner}}\ and\ \bibinfo {author} {\bibfnamefont {P.}~\bibnamefont
  {Zoller}},\ }\href@noop {} {\bibfield  {journal} {\bibinfo  {journal}
  {Physical Review A}\ }\textbf {\bibinfo {volume} {66}},\ \bibinfo {pages}
  {023816} (\bibinfo {year} {2002})}\BibitemShut {NoStop}%
\bibitem [{\citenamefont {K{\"a}stel}\ and\ \citenamefont
  {Fleischhauer}(2005)}]{kastel2005suppression}%
  \BibitemOpen
  \bibfield  {author} {\bibinfo {author} {\bibfnamefont {J.}~\bibnamefont
  {K{\"a}stel}}\ and\ \bibinfo {author} {\bibfnamefont {M.}~\bibnamefont
  {Fleischhauer}},\ }\href@noop {} {\bibfield  {journal} {\bibinfo  {journal}
  {Physical Review A}\ }\textbf {\bibinfo {volume} {71}},\ \bibinfo {pages}
  {011804} (\bibinfo {year} {2005})}\BibitemShut {NoStop}%
\bibitem [{\citenamefont {Eschner}\ \emph {et~al.}(2001)\citenamefont
  {Eschner}, \citenamefont {Raab}, \citenamefont {Schmidt-Kaler},\ and\
  \citenamefont {Blatt}}]{eschner2001light}%
  \BibitemOpen
  \bibfield  {author} {\bibinfo {author} {\bibfnamefont {J.}~\bibnamefont
  {Eschner}}, \bibinfo {author} {\bibfnamefont {C.}~\bibnamefont {Raab}},
  \bibinfo {author} {\bibfnamefont {F.}~\bibnamefont {Schmidt-Kaler}}, \ and\
  \bibinfo {author} {\bibfnamefont {R.}~\bibnamefont {Blatt}},\ }\href@noop {}
  {\bibfield  {journal} {\bibinfo  {journal} {Nature}\ }\textbf {\bibinfo
  {volume} {413}},\ \bibinfo {pages} {495} (\bibinfo {year}
  {2001})}\BibitemShut {NoStop}%
\bibitem [{\citenamefont {Jha}\ \emph {et~al.}(2015)\citenamefont {Jha},
  \citenamefont {Ni}, \citenamefont {Wu}, \citenamefont {Wang},\ and\
  \citenamefont {Zhang}}]{jha2015metasurface}%
  \BibitemOpen
  \bibfield  {author} {\bibinfo {author} {\bibfnamefont {P.~K.}\ \bibnamefont
  {Jha}}, \bibinfo {author} {\bibfnamefont {X.}~\bibnamefont {Ni}}, \bibinfo
  {author} {\bibfnamefont {C.}~\bibnamefont {Wu}}, \bibinfo {author}
  {\bibfnamefont {Y.}~\bibnamefont {Wang}}, \ and\ \bibinfo {author}
  {\bibfnamefont {X.}~\bibnamefont {Zhang}},\ }\href@noop {} {\bibfield
  {journal} {\bibinfo  {journal} {Physical Review Letters}\ }\textbf {\bibinfo
  {volume} {115}},\ \bibinfo {pages} {025501} (\bibinfo {year}
  {2015})}\BibitemShut {NoStop}%
\bibitem [{\citenamefont {Jha}\ \emph {et~al.}(2018)\citenamefont {Jha},
  \citenamefont {Shitrit}, \citenamefont {Ren}, \citenamefont {Wang},\ and\
  \citenamefont {Zhang}}]{jha2018spontaneous}%
  \BibitemOpen
  \bibfield  {author} {\bibinfo {author} {\bibfnamefont {P.~K.}\ \bibnamefont
  {Jha}}, \bibinfo {author} {\bibfnamefont {N.}~\bibnamefont {Shitrit}},
  \bibinfo {author} {\bibfnamefont {X.}~\bibnamefont {Ren}}, \bibinfo {author}
  {\bibfnamefont {Y.}~\bibnamefont {Wang}}, \ and\ \bibinfo {author}
  {\bibfnamefont {X.}~\bibnamefont {Zhang}},\ }\href@noop {} {\bibfield
  {journal} {\bibinfo  {journal} {Physical Review Letters}\ }\textbf {\bibinfo
  {volume} {121}},\ \bibinfo {pages} {116102} (\bibinfo {year}
  {2018})}\BibitemShut {NoStop}%
\bibitem [{\citenamefont {Lalanne}\ and\ \citenamefont
  {Chavel}(2017)}]{lalanne2017metalenses}%
  \BibitemOpen
  \bibfield  {author} {\bibinfo {author} {\bibfnamefont {P.}~\bibnamefont
  {Lalanne}}\ and\ \bibinfo {author} {\bibfnamefont {P.}~\bibnamefont
  {Chavel}},\ }\href@noop {} {\bibfield  {journal} {\bibinfo  {journal} {Laser
  \& Photonics Reviews}\ }\textbf {\bibinfo {volume} {11}},\ \bibinfo {pages}
  {1600295} (\bibinfo {year} {2017})}\BibitemShut {NoStop}%
\bibitem [{\citenamefont {Genevet}\ \emph {et~al.}(2017)\citenamefont
  {Genevet}, \citenamefont {Capasso}, \citenamefont {Aieta}, \citenamefont
  {Khorasaninejad},\ and\ \citenamefont {Devlin}}]{genevet2017recent}%
  \BibitemOpen
  \bibfield  {author} {\bibinfo {author} {\bibfnamefont {P.}~\bibnamefont
  {Genevet}}, \bibinfo {author} {\bibfnamefont {F.}~\bibnamefont {Capasso}},
  \bibinfo {author} {\bibfnamefont {F.}~\bibnamefont {Aieta}}, \bibinfo
  {author} {\bibfnamefont {M.}~\bibnamefont {Khorasaninejad}}, \ and\ \bibinfo
  {author} {\bibfnamefont {R.}~\bibnamefont {Devlin}},\ }\href@noop {}
  {\bibfield  {journal} {\bibinfo  {journal} {Optica}\ }\textbf {\bibinfo
  {volume} {4}},\ \bibinfo {pages} {139} (\bibinfo {year} {2017})}\BibitemShut
  {NoStop}%
\bibitem [{\citenamefont {Agarwal}(2000)}]{agarwal2000anisotropic}%
  \BibitemOpen
  \bibfield  {author} {\bibinfo {author} {\bibfnamefont {G.}~\bibnamefont
  {Agarwal}},\ }\href@noop {} {\bibfield  {journal} {\bibinfo  {journal}
  {Physical Review Letters}\ }\textbf {\bibinfo {volume} {84}},\ \bibinfo
  {pages} {5500} (\bibinfo {year} {2000})}\BibitemShut {NoStop}%
\bibitem [{\citenamefont {Agarwal}\ and\ \citenamefont
  {Patnaik}(2001)}]{agarwal2001vacuum}%
  \BibitemOpen
  \bibfield  {author} {\bibinfo {author} {\bibfnamefont {G.}~\bibnamefont
  {Agarwal}}\ and\ \bibinfo {author} {\bibfnamefont {A.~K.}\ \bibnamefont
  {Patnaik}},\ }\href@noop {} {\bibfield  {journal} {\bibinfo  {journal}
  {Physical Review A}\ }\textbf {\bibinfo {volume} {63}},\ \bibinfo {pages}
  {043805} (\bibinfo {year} {2001})}\BibitemShut {NoStop}%
\bibitem [{\citenamefont {Li}\ \emph {et~al.}(2001)\citenamefont {Li},
  \citenamefont {Li},\ and\ \citenamefont {Zhu}}]{li2001quantum}%
  \BibitemOpen
  \bibfield  {author} {\bibinfo {author} {\bibfnamefont {G.-X.}\ \bibnamefont
  {Li}}, \bibinfo {author} {\bibfnamefont {F.-L.}\ \bibnamefont {Li}}, \ and\
  \bibinfo {author} {\bibfnamefont {S.-Y.}\ \bibnamefont {Zhu}},\ }\href@noop
  {} {\bibfield  {journal} {\bibinfo  {journal} {Physical Review A}\ }\textbf
  {\bibinfo {volume} {64}},\ \bibinfo {pages} {013819} (\bibinfo {year}
  {2001})}\BibitemShut {NoStop}%
\bibitem [{\citenamefont {Yang}\ \emph {et~al.}(2008)\citenamefont {Yang},
  \citenamefont {Xu}, \citenamefont {Chen},\ and\ \citenamefont
  {Zhu}}]{yang2008quantum}%
  \BibitemOpen
  \bibfield  {author} {\bibinfo {author} {\bibfnamefont {Y.}~\bibnamefont
  {Yang}}, \bibinfo {author} {\bibfnamefont {J.}~\bibnamefont {Xu}}, \bibinfo
  {author} {\bibfnamefont {H.}~\bibnamefont {Chen}}, \ and\ \bibinfo {author}
  {\bibfnamefont {S.}~\bibnamefont {Zhu}},\ }\href@noop {} {\bibfield
  {journal} {\bibinfo  {journal} {Physical Review Letters}\ }\textbf {\bibinfo
  {volume} {100}},\ \bibinfo {pages} {043601} (\bibinfo {year}
  {2008})}\BibitemShut {NoStop}%
\bibitem [{\citenamefont {Sun}\ and\ \citenamefont
  {Jiang}(2016)}]{sun2016quantum}%
  \BibitemOpen
  \bibfield  {author} {\bibinfo {author} {\bibfnamefont {L.}~\bibnamefont
  {Sun}}\ and\ \bibinfo {author} {\bibfnamefont {C.}~\bibnamefont {Jiang}},\
  }\href@noop {} {\bibfield  {journal} {\bibinfo  {journal} {Optics Express}\
  }\textbf {\bibinfo {volume} {24}},\ \bibinfo {pages} {7719} (\bibinfo {year}
  {2016})}\BibitemShut {NoStop}%
\bibitem [{\citenamefont {Hughes}\ and\ \citenamefont
  {Agarwal}(2017)}]{hughes2017anisotropy}%
  \BibitemOpen
  \bibfield  {author} {\bibinfo {author} {\bibfnamefont {S.}~\bibnamefont
  {Hughes}}\ and\ \bibinfo {author} {\bibfnamefont {G.~S.}\ \bibnamefont
  {Agarwal}},\ }\href@noop {} {\bibfield  {journal} {\bibinfo  {journal}
  {Physical Review Letters}\ }\textbf {\bibinfo {volume} {118}},\ \bibinfo
  {pages} {063601} (\bibinfo {year} {2017})}\BibitemShut {NoStop}%
\bibitem [{\citenamefont {Suter}(1997)}]{suter1997physics}%
  \BibitemOpen
  \bibfield  {author} {\bibinfo {author} {\bibfnamefont {D.}~\bibnamefont
  {Suter}},\ }\href@noop {} {\emph {\bibinfo {title} {The physics of laser-atom
  interactions}}},\ Vol.~\bibinfo {volume} {19}\ (\bibinfo  {publisher}
  {Cambridge University Press},\ \bibinfo {year} {1997})\BibitemShut {NoStop}%
\bibitem [{\citenamefont {Togan}\ \emph {et~al.}(2010)\citenamefont {Togan},
  \citenamefont {Chu}, \citenamefont {Trifonov}, \citenamefont {Jiang},
  \citenamefont {Maze}, \citenamefont {Childress}, \citenamefont {Dutt},
  \citenamefont {S{\o}rensen}, \citenamefont {Hemmer}, \citenamefont {Zibrov}
  \emph {et~al.}}]{togan2010quantum}%
  \BibitemOpen
  \bibfield  {author} {\bibinfo {author} {\bibfnamefont {E.}~\bibnamefont
  {Togan}}, \bibinfo {author} {\bibfnamefont {Y.}~\bibnamefont {Chu}}, \bibinfo
  {author} {\bibfnamefont {A.}~\bibnamefont {Trifonov}}, \bibinfo {author}
  {\bibfnamefont {L.}~\bibnamefont {Jiang}}, \bibinfo {author} {\bibfnamefont
  {J.}~\bibnamefont {Maze}}, \bibinfo {author} {\bibfnamefont {L.}~\bibnamefont
  {Childress}}, \bibinfo {author} {\bibfnamefont {M.~G.}\ \bibnamefont {Dutt}},
  \bibinfo {author} {\bibfnamefont {A.~S.}\ \bibnamefont {S{\o}rensen}},
  \bibinfo {author} {\bibfnamefont {P.}~\bibnamefont {Hemmer}}, \bibinfo
  {author} {\bibfnamefont {A.~S.}\ \bibnamefont {Zibrov}},  \emph {et~al.},\
  }\href@noop {} {\bibfield  {journal} {\bibinfo  {journal} {Nature}\ }\textbf
  {\bibinfo {volume} {466}},\ \bibinfo {pages} {730} (\bibinfo {year}
  {2010})}\BibitemShut {NoStop}%
\bibitem [{\citenamefont {Barnett}\ and\ \citenamefont
  {Radmore}(2002)}]{barnett2002methods}%
  \BibitemOpen
  \bibfield  {author} {\bibinfo {author} {\bibfnamefont {S.~M.}\ \bibnamefont
  {Barnett}}\ and\ \bibinfo {author} {\bibfnamefont {P.~M.}\ \bibnamefont
  {Radmore}},\ }\href@noop {} {\emph {\bibinfo {title} {Methods in theoretical
  quantum optics}}},\ Vol.~\bibinfo {volume} {15}\ (\bibinfo  {publisher}
  {Oxford University Press},\ \bibinfo {year} {2002})\BibitemShut {NoStop}%
\bibitem [{\citenamefont {Carmichael}(2013)}]{carmichael2013statistical}%
  \BibitemOpen
  \bibfield  {author} {\bibinfo {author} {\bibfnamefont {H.~J.}\ \bibnamefont
  {Carmichael}},\ }\href@noop {} {\emph {\bibinfo {title} {Statistical methods
  in quantum optics 1: master equations and Fokker-Planck equations}}}\
  (\bibinfo  {publisher} {Springer Science \& Business Media},\ \bibinfo {year}
  {2013})\BibitemShut {NoStop}%
\bibitem [{\citenamefont {Balasubramanian}\ \emph {et~al.}(2009)\citenamefont
  {Balasubramanian}, \citenamefont {Neumann}, \citenamefont {Twitchen},
  \citenamefont {Markham}, \citenamefont {Kolesov}, \citenamefont {Mizuochi},
  \citenamefont {Isoya}, \citenamefont {Achard}, \citenamefont {Beck},
  \citenamefont {Tissler}, \citenamefont {Jacques}, \citenamefont {Hemmer},
  \citenamefont {Jelezko},\ and\ \citenamefont
  {Wrachtrup}}]{balasubramanian_ultralong_2009}%
  \BibitemOpen
  \bibfield  {author} {\bibinfo {author} {\bibfnamefont {G.}~\bibnamefont
  {Balasubramanian}}, \bibinfo {author} {\bibfnamefont {P.}~\bibnamefont
  {Neumann}}, \bibinfo {author} {\bibfnamefont {D.}~\bibnamefont {Twitchen}},
  \bibinfo {author} {\bibfnamefont {M.}~\bibnamefont {Markham}}, \bibinfo
  {author} {\bibfnamefont {R.}~\bibnamefont {Kolesov}}, \bibinfo {author}
  {\bibfnamefont {N.}~\bibnamefont {Mizuochi}}, \bibinfo {author}
  {\bibfnamefont {J.}~\bibnamefont {Isoya}}, \bibinfo {author} {\bibfnamefont
  {J.}~\bibnamefont {Achard}}, \bibinfo {author} {\bibfnamefont
  {J.}~\bibnamefont {Beck}}, \bibinfo {author} {\bibfnamefont {J.}~\bibnamefont
  {Tissler}}, \bibinfo {author} {\bibfnamefont {V.}~\bibnamefont {Jacques}},
  \bibinfo {author} {\bibfnamefont {P.~R.}\ \bibnamefont {Hemmer}}, \bibinfo
  {author} {\bibfnamefont {F.}~\bibnamefont {Jelezko}}, \ and\ \bibinfo
  {author} {\bibfnamefont {J.}~\bibnamefont {Wrachtrup}},\ }\href {\doibase
  10.1038/nmat2420} {\bibfield  {journal} {\bibinfo  {journal} {Nature
  Materials}\ }\textbf {\bibinfo {volume} {8}},\ \bibinfo {pages} {383}
  (\bibinfo {year} {2009})}\BibitemShut {NoStop}%
\bibitem [{\citenamefont {Scully}\ and\ \citenamefont
  {Zubairy}(1997)}]{scully1999quantum}%
  \BibitemOpen
  \bibfield  {author} {\bibinfo {author} {\bibfnamefont {M.~O.}\ \bibnamefont
  {Scully}}\ and\ \bibinfo {author} {\bibfnamefont {M.~S.}\ \bibnamefont
  {Zubairy}},\ }\href@noop {} {\emph {\bibinfo {title} {Quantum optics}}}\
  (\bibinfo  {publisher} {Cambridge University Press},\ \bibinfo {year}
  {1997})\BibitemShut {NoStop}%
\bibitem [{\citenamefont {Jaeger}\ \emph {et~al.}(1995)\citenamefont {Jaeger},
  \citenamefont {Shimony},\ and\ \citenamefont {Vaidman}}]{jaeger1995two}%
  \BibitemOpen
  \bibfield  {author} {\bibinfo {author} {\bibfnamefont {G.}~\bibnamefont
  {Jaeger}}, \bibinfo {author} {\bibfnamefont {A.}~\bibnamefont {Shimony}}, \
  and\ \bibinfo {author} {\bibfnamefont {L.}~\bibnamefont {Vaidman}},\
  }\href@noop {} {\bibfield  {journal} {\bibinfo  {journal} {Physical Review
  A}\ }\textbf {\bibinfo {volume} {51}},\ \bibinfo {pages} {54} (\bibinfo
  {year} {1995})}\BibitemShut {NoStop}%
\bibitem [{\citenamefont {Englert}(1996)}]{englert1996fringe}%
  \BibitemOpen
  \bibfield  {author} {\bibinfo {author} {\bibfnamefont {B.-G.}\ \bibnamefont
  {Englert}},\ }\href@noop {} {\bibfield  {journal} {\bibinfo  {journal}
  {Physical Review Letters}\ }\textbf {\bibinfo {volume} {77}},\ \bibinfo
  {pages} {2154} (\bibinfo {year} {1996})}\BibitemShut {NoStop}%
\bibitem [{\citenamefont {Yannopapas}\ \emph {et~al.}(2009)\citenamefont
  {Yannopapas}, \citenamefont {Paspalakis},\ and\ \citenamefont
  {Vitanov}}]{PhysRevLett.103.063602}%
  \BibitemOpen
  \bibfield  {author} {\bibinfo {author} {\bibfnamefont {V.}~\bibnamefont
  {Yannopapas}}, \bibinfo {author} {\bibfnamefont {E.}~\bibnamefont
  {Paspalakis}}, \ and\ \bibinfo {author} {\bibfnamefont {N.~V.}\ \bibnamefont
  {Vitanov}},\ }\href {\doibase 10.1103/PhysRevLett.103.063602} {\bibfield
  {journal} {\bibinfo  {journal} {Phys. Rev. Lett.}\ }\textbf {\bibinfo
  {volume} {103}},\ \bibinfo {pages} {063602} (\bibinfo {year}
  {2009})}\BibitemShut {NoStop}%
\bibitem [{\citenamefont {Jhe}\ \emph {et~al.}(1987)\citenamefont {Jhe},
  \citenamefont {Anderson}, \citenamefont {Hinds}, \citenamefont {Meschede},
  \citenamefont {Moi},\ and\ \citenamefont {Haroche}}]{jhe1987suppression}%
  \BibitemOpen
  \bibfield  {author} {\bibinfo {author} {\bibfnamefont {W.}~\bibnamefont
  {Jhe}}, \bibinfo {author} {\bibfnamefont {A.}~\bibnamefont {Anderson}},
  \bibinfo {author} {\bibfnamefont {E.}~\bibnamefont {Hinds}}, \bibinfo
  {author} {\bibfnamefont {D.}~\bibnamefont {Meschede}}, \bibinfo {author}
  {\bibfnamefont {L.}~\bibnamefont {Moi}}, \ and\ \bibinfo {author}
  {\bibfnamefont {S.}~\bibnamefont {Haroche}},\ }\href@noop {} {\bibfield
  {journal} {\bibinfo  {journal} {Physical Review Letters}\ }\textbf {\bibinfo
  {volume} {58}},\ \bibinfo {pages} {666} (\bibinfo {year} {1987})}\BibitemShut
  {NoStop}%
\bibitem [{\citenamefont {Taillandier-Loize}\ \emph {et~al.}(2014)\citenamefont
  {Taillandier-Loize}, \citenamefont {Baudon}, \citenamefont {Dutier},
  \citenamefont {Perales}, \citenamefont {Boustimi},\ and\ \citenamefont
  {Ducloy}}]{taillandier2014anisotropic}%
  \BibitemOpen
  \bibfield  {author} {\bibinfo {author} {\bibfnamefont {T.}~\bibnamefont
  {Taillandier-Loize}}, \bibinfo {author} {\bibfnamefont {J.}~\bibnamefont
  {Baudon}}, \bibinfo {author} {\bibfnamefont {G.}~\bibnamefont {Dutier}},
  \bibinfo {author} {\bibfnamefont {F.}~\bibnamefont {Perales}}, \bibinfo
  {author} {\bibfnamefont {M.}~\bibnamefont {Boustimi}}, \ and\ \bibinfo
  {author} {\bibfnamefont {M.}~\bibnamefont {Ducloy}},\ }\href@noop {}
  {\bibfield  {journal} {\bibinfo  {journal} {Physical Review A}\ }\textbf
  {\bibinfo {volume} {89}},\ \bibinfo {pages} {052514} (\bibinfo {year}
  {2014})}\BibitemShut {NoStop}%
\bibitem [{\citenamefont {Boustimi}\ \emph {et~al.}(2001)\citenamefont
  {Boustimi}, \citenamefont {De~Lesegno}, \citenamefont {Baudon}, \citenamefont
  {Robert},\ and\ \citenamefont {Ducloy}}]{boustimi2001atom}%
  \BibitemOpen
  \bibfield  {author} {\bibinfo {author} {\bibfnamefont {M.}~\bibnamefont
  {Boustimi}}, \bibinfo {author} {\bibfnamefont {B.~V.}\ \bibnamefont
  {De~Lesegno}}, \bibinfo {author} {\bibfnamefont {J.}~\bibnamefont {Baudon}},
  \bibinfo {author} {\bibfnamefont {J.}~\bibnamefont {Robert}}, \ and\ \bibinfo
  {author} {\bibfnamefont {M.}~\bibnamefont {Ducloy}},\ }\href@noop {}
  {\bibfield  {journal} {\bibinfo  {journal} {Physical Review Letters}\
  }\textbf {\bibinfo {volume} {86}},\ \bibinfo {pages} {2766} (\bibinfo {year}
  {2001})}\BibitemShut {NoStop}%
\bibitem [{\citenamefont {Thanopulos}\ \emph {et~al.}(2019)\citenamefont
  {Thanopulos}, \citenamefont {Karanikolas},\ and\ \citenamefont
  {Paspalakis}}]{Thanopulos:19}%
  \BibitemOpen
  \bibfield  {author} {\bibinfo {author} {\bibfnamefont {I.}~\bibnamefont
  {Thanopulos}}, \bibinfo {author} {\bibfnamefont {V.}~\bibnamefont
  {Karanikolas}}, \ and\ \bibinfo {author} {\bibfnamefont {E.}~\bibnamefont
  {Paspalakis}},\ }\href {\doibase 10.1364/OL.44.003510} {\bibfield  {journal}
  {\bibinfo  {journal} {Opt. Lett.}\ }\textbf {\bibinfo {volume} {44}},\
  \bibinfo {pages} {3510} (\bibinfo {year} {2019})}\BibitemShut {NoStop}%
\bibitem [{\citenamefont {Evangelou}\ \emph {et~al.}(2011)\citenamefont
  {Evangelou}, \citenamefont {Yannopapas},\ and\ \citenamefont
  {Paspalakis}}]{PhysRevA.83.055805}%
  \BibitemOpen
  \bibfield  {author} {\bibinfo {author} {\bibfnamefont {S.}~\bibnamefont
  {Evangelou}}, \bibinfo {author} {\bibfnamefont {V.}~\bibnamefont
  {Yannopapas}}, \ and\ \bibinfo {author} {\bibfnamefont {E.}~\bibnamefont
  {Paspalakis}},\ }\href {\doibase 10.1103/PhysRevA.83.055805} {\bibfield
  {journal} {\bibinfo  {journal} {Phys. Rev. A}\ }\textbf {\bibinfo {volume}
  {83}},\ \bibinfo {pages} {055805} (\bibinfo {year} {2011})}\BibitemShut
  {NoStop}%
\bibitem [{\citenamefont {Thanopulos}\ \emph {et~al.}(2017)\citenamefont
  {Thanopulos}, \citenamefont {Yannopapas},\ and\ \citenamefont
  {Paspalakis}}]{PhysRevB.95.075412}%
  \BibitemOpen
  \bibfield  {author} {\bibinfo {author} {\bibfnamefont {I.}~\bibnamefont
  {Thanopulos}}, \bibinfo {author} {\bibfnamefont {V.}~\bibnamefont
  {Yannopapas}}, \ and\ \bibinfo {author} {\bibfnamefont {E.}~\bibnamefont
  {Paspalakis}},\ }\href {\doibase 10.1103/PhysRevB.95.075412} {\bibfield
  {journal} {\bibinfo  {journal} {Phys. Rev. B}\ }\textbf {\bibinfo {volume}
  {95}},\ \bibinfo {pages} {075412} (\bibinfo {year} {2017})}\BibitemShut
  {NoStop}%
\bibitem [{\citenamefont {Karanikolas}\ and\ \citenamefont
  {Paspalakis}(2018)}]{doi:10.1021/acs.jpcc.8b02703}%
  \BibitemOpen
  \bibfield  {author} {\bibinfo {author} {\bibfnamefont {V.}~\bibnamefont
  {Karanikolas}}\ and\ \bibinfo {author} {\bibfnamefont {E.}~\bibnamefont
  {Paspalakis}},\ }\href {\doibase 10.1021/acs.jpcc.8b02703} {\bibfield
  {journal} {\bibinfo  {journal} {The Journal of Physical Chemistry C}\
  }\textbf {\bibinfo {volume} {122}},\ \bibinfo {pages} {14788} (\bibinfo
  {year} {2018})}\BibitemShut {NoStop}%
\bibitem [{\citenamefont {Sun}\ \emph {et~al.}(2012)\citenamefont {Sun},
  \citenamefont {Yang}, \citenamefont {Wang}, \citenamefont {Juan},
  \citenamefont {Chen}, \citenamefont {Liao}, \citenamefont {He}, \citenamefont
  {Xiao}, \citenamefont {Kung}, \citenamefont {Guo} \emph
  {et~al.}}]{sun2012high}%
  \BibitemOpen
  \bibfield  {author} {\bibinfo {author} {\bibfnamefont {S.}~\bibnamefont
  {Sun}}, \bibinfo {author} {\bibfnamefont {K.-Y.}\ \bibnamefont {Yang}},
  \bibinfo {author} {\bibfnamefont {C.-M.}\ \bibnamefont {Wang}}, \bibinfo
  {author} {\bibfnamefont {T.-K.}\ \bibnamefont {Juan}}, \bibinfo {author}
  {\bibfnamefont {W.~T.}\ \bibnamefont {Chen}}, \bibinfo {author}
  {\bibfnamefont {C.~Y.}\ \bibnamefont {Liao}}, \bibinfo {author}
  {\bibfnamefont {Q.}~\bibnamefont {He}}, \bibinfo {author} {\bibfnamefont
  {S.}~\bibnamefont {Xiao}}, \bibinfo {author} {\bibfnamefont {W.-T.}\
  \bibnamefont {Kung}}, \bibinfo {author} {\bibfnamefont {G.-Y.}\ \bibnamefont
  {Guo}},  \emph {et~al.},\ }\href@noop {} {\bibfield  {journal} {\bibinfo
  {journal} {Nano Letters}\ }\textbf {\bibinfo {volume} {12}},\ \bibinfo
  {pages} {6223} (\bibinfo {year} {2012})}\BibitemShut {NoStop}%
\bibitem [{\citenamefont {Pors}\ \emph {et~al.}(2013)\citenamefont {Pors},
  \citenamefont {Albrektsen}, \citenamefont {Radko},\ and\ \citenamefont
  {Bozhevolnyi}}]{pors2013gap}%
  \BibitemOpen
  \bibfield  {author} {\bibinfo {author} {\bibfnamefont {A.}~\bibnamefont
  {Pors}}, \bibinfo {author} {\bibfnamefont {O.}~\bibnamefont {Albrektsen}},
  \bibinfo {author} {\bibfnamefont {I.~P.}\ \bibnamefont {Radko}}, \ and\
  \bibinfo {author} {\bibfnamefont {S.~I.}\ \bibnamefont {Bozhevolnyi}},\
  }\href@noop {} {\bibfield  {journal} {\bibinfo  {journal} {Scientific
  Reports}\ }\textbf {\bibinfo {volume} {3}},\ \bibinfo {pages} {2155}
  (\bibinfo {year} {2013})}\BibitemShut {NoStop}%
\bibitem [{\citenamefont {Zheng}\ \emph {et~al.}(2015)\citenamefont {Zheng},
  \citenamefont {M{\"u}hlenbernd}, \citenamefont {Kenney}, \citenamefont {Li},
  \citenamefont {Zentgraf},\ and\ \citenamefont
  {Zhang}}]{zheng2015metasurface}%
  \BibitemOpen
  \bibfield  {author} {\bibinfo {author} {\bibfnamefont {G.}~\bibnamefont
  {Zheng}}, \bibinfo {author} {\bibfnamefont {H.}~\bibnamefont
  {M{\"u}hlenbernd}}, \bibinfo {author} {\bibfnamefont {M.}~\bibnamefont
  {Kenney}}, \bibinfo {author} {\bibfnamefont {G.}~\bibnamefont {Li}}, \bibinfo
  {author} {\bibfnamefont {T.}~\bibnamefont {Zentgraf}}, \ and\ \bibinfo
  {author} {\bibfnamefont {S.}~\bibnamefont {Zhang}},\ }\href@noop {}
  {\bibfield  {journal} {\bibinfo  {journal} {Nature Nanotechnology}\ }\textbf
  {\bibinfo {volume} {10}},\ \bibinfo {pages} {308} (\bibinfo {year}
  {2015})}\BibitemShut {NoStop}%
\bibitem [{\citenamefont {Jha}\ \emph {et~al.}(2017)\citenamefont {Jha},
  \citenamefont {Shitrit}, \citenamefont {Kim}, \citenamefont {Ren},
  \citenamefont {Wang},\ and\ \citenamefont {Zhang}}]{jha2017metasurface}%
  \BibitemOpen
  \bibfield  {author} {\bibinfo {author} {\bibfnamefont {P.~K.}\ \bibnamefont
  {Jha}}, \bibinfo {author} {\bibfnamefont {N.}~\bibnamefont {Shitrit}},
  \bibinfo {author} {\bibfnamefont {J.}~\bibnamefont {Kim}}, \bibinfo {author}
  {\bibfnamefont {X.}~\bibnamefont {Ren}}, \bibinfo {author} {\bibfnamefont
  {Y.}~\bibnamefont {Wang}}, \ and\ \bibinfo {author} {\bibfnamefont
  {X.}~\bibnamefont {Zhang}},\ }\href@noop {} {\bibfield  {journal} {\bibinfo
  {journal} {ACS Photonics}\ }\textbf {\bibinfo {volume} {5}},\ \bibinfo
  {pages} {971} (\bibinfo {year} {2017})}\BibitemShut {NoStop}%
\bibitem [{\citenamefont {Hugonin}\ and\ \citenamefont
  {Lalanne}(2005)}]{hugonin2005reticolo}%
  \BibitemOpen
  \bibfield  {author} {\bibinfo {author} {\bibfnamefont {J.}~\bibnamefont
  {Hugonin}}\ and\ \bibinfo {author} {\bibfnamefont {P.}~\bibnamefont
  {Lalanne}},\ }\href@noop {} {\emph {\bibinfo {title} {Reticolo software for
  grating analysis}}}\ (\bibinfo  {publisher} {Institute d'Optique, Palaiseau,
  France},\ \bibinfo {year} {2005})\BibitemShut {NoStop}%
\bibitem [{\citenamefont {Moharam}\ \emph {et~al.}(1995)\citenamefont
  {Moharam}, \citenamefont {Grann}, \citenamefont {Pommet},\ and\ \citenamefont
  {Gaylord}}]{moharam1995formulation}%
  \BibitemOpen
  \bibfield  {author} {\bibinfo {author} {\bibfnamefont {M.}~\bibnamefont
  {Moharam}}, \bibinfo {author} {\bibfnamefont {E.~B.}\ \bibnamefont {Grann}},
  \bibinfo {author} {\bibfnamefont {D.~A.}\ \bibnamefont {Pommet}}, \ and\
  \bibinfo {author} {\bibfnamefont {T.}~\bibnamefont {Gaylord}},\ }\href@noop
  {} {\bibfield  {journal} {\bibinfo  {journal} {JOSA A}\ }\textbf {\bibinfo
  {volume} {12}},\ \bibinfo {pages} {1068} (\bibinfo {year}
  {1995})}\BibitemShut {NoStop}%
\bibitem [{\citenamefont {Li}(1997)}]{li1997new}%
  \BibitemOpen
  \bibfield  {author} {\bibinfo {author} {\bibfnamefont {L.}~\bibnamefont
  {Li}},\ }\href@noop {} {\bibfield  {journal} {\bibinfo  {journal} {JOSA A}\
  }\textbf {\bibinfo {volume} {14}},\ \bibinfo {pages} {2758} (\bibinfo {year}
  {1997})}\BibitemShut {NoStop}%
\bibitem [{\citenamefont {Lalanne}\ and\ \citenamefont
  {Jurek}(1998)}]{lalanne1998computation}%
  \BibitemOpen
  \bibfield  {author} {\bibinfo {author} {\bibfnamefont {P.}~\bibnamefont
  {Lalanne}}\ and\ \bibinfo {author} {\bibfnamefont {M.~P.}\ \bibnamefont
  {Jurek}},\ }\href@noop {} {\bibfield  {journal} {\bibinfo  {journal} {Journal
  of Modern Optics}\ }\textbf {\bibinfo {volume} {45}},\ \bibinfo {pages}
  {1357} (\bibinfo {year} {1998})}\BibitemShut {NoStop}%
\bibitem [{\citenamefont {Popov}\ and\ \citenamefont
  {Nevi{\`e}re}(2000)}]{popov2000grating}%
  \BibitemOpen
  \bibfield  {author} {\bibinfo {author} {\bibfnamefont {E.}~\bibnamefont
  {Popov}}\ and\ \bibinfo {author} {\bibfnamefont {M.}~\bibnamefont
  {Nevi{\`e}re}},\ }\href@noop {} {\bibfield  {journal} {\bibinfo  {journal}
  {JOSA A}\ }\textbf {\bibinfo {volume} {17}},\ \bibinfo {pages} {1773}
  (\bibinfo {year} {2000})}\BibitemShut {NoStop}%
\bibitem [{\citenamefont {Luo}\ \emph {et~al.}(2015)\citenamefont {Luo},
  \citenamefont {Xiao}, \citenamefont {He}, \citenamefont {Sun},\ and\
  \citenamefont {Zhou}}]{luo2015photonic}%
  \BibitemOpen
  \bibfield  {author} {\bibinfo {author} {\bibfnamefont {W.}~\bibnamefont
  {Luo}}, \bibinfo {author} {\bibfnamefont {S.}~\bibnamefont {Xiao}}, \bibinfo
  {author} {\bibfnamefont {Q.}~\bibnamefont {He}}, \bibinfo {author}
  {\bibfnamefont {S.}~\bibnamefont {Sun}}, \ and\ \bibinfo {author}
  {\bibfnamefont {L.}~\bibnamefont {Zhou}},\ }\href@noop {} {\bibfield
  {journal} {\bibinfo  {journal} {Advanced Optical Materials}\ }\textbf
  {\bibinfo {volume} {3}},\ \bibinfo {pages} {1102} (\bibinfo {year}
  {2015})}\BibitemShut {NoStop}%
\bibitem [{\citenamefont {Larouche}\ and\ \citenamefont
  {Smith}(2012)}]{Larouche:12}%
  \BibitemOpen
  \bibfield  {author} {\bibinfo {author} {\bibfnamefont {S.}~\bibnamefont
  {Larouche}}\ and\ \bibinfo {author} {\bibfnamefont {D.~R.}\ \bibnamefont
  {Smith}},\ }\href {\doibase 10.1364/OL.37.002391} {\bibfield  {journal}
  {\bibinfo  {journal} {Optics Letters}\ }\textbf {\bibinfo {volume} {37}},\
  \bibinfo {pages} {2391} (\bibinfo {year} {2012})}\BibitemShut {NoStop}%
\bibitem [{\citenamefont {Pancharatnam}(1956)}]{pancharatnam1956generalized}%
  \BibitemOpen
  \bibfield  {author} {\bibinfo {author} {\bibfnamefont {S.}~\bibnamefont
  {Pancharatnam}},\ }in\ \href@noop {} {\emph {\bibinfo {booktitle}
  {Proceedings of the Indian Academy of Sciences-Section A}}},\ Vol.~\bibinfo
  {volume} {44}\ (\bibinfo {organization} {Springer},\ \bibinfo {year} {1956})\
  pp.\ \bibinfo {pages} {398--417}\BibitemShut {NoStop}%
\bibitem [{\citenamefont {Berry}(1987)}]{berry1987adiabatic}%
  \BibitemOpen
  \bibfield  {author} {\bibinfo {author} {\bibfnamefont {M.~V.}\ \bibnamefont
  {Berry}},\ }\href@noop {} {\bibfield  {journal} {\bibinfo  {journal} {Journal
  of Modern Optics}\ }\textbf {\bibinfo {volume} {34}},\ \bibinfo {pages}
  {1401} (\bibinfo {year} {1987})}\BibitemShut {NoStop}%
\bibitem [{\citenamefont {Lassalle}\ \emph {et~al.}(2018)\citenamefont
  {Lassalle}, \citenamefont {Bonod}, \citenamefont {Durt},\ and\ \citenamefont
  {Stout}}]{lassalle2018interplay}%
  \BibitemOpen
  \bibfield  {author} {\bibinfo {author} {\bibfnamefont {E.}~\bibnamefont
  {Lassalle}}, \bibinfo {author} {\bibfnamefont {N.}~\bibnamefont {Bonod}},
  \bibinfo {author} {\bibfnamefont {T.}~\bibnamefont {Durt}}, \ and\ \bibinfo
  {author} {\bibfnamefont {B.}~\bibnamefont {Stout}},\ }\href@noop {}
  {\bibfield  {journal} {\bibinfo  {journal} {Optics Letters}\ }\textbf
  {\bibinfo {volume} {43}},\ \bibinfo {pages} {1950} (\bibinfo {year}
  {2018})}\BibitemShut {NoStop}%
\bibitem [{\citenamefont {Ong}\ \emph {et~al.}(2017)\citenamefont {Ong},
  \citenamefont {Chu}, \citenamefont {Chen}, \citenamefont {Zhu},\ and\
  \citenamefont {Genevet}}]{ong2017freestanding}%
  \BibitemOpen
  \bibfield  {author} {\bibinfo {author} {\bibfnamefont {J.~R.}\ \bibnamefont
  {Ong}}, \bibinfo {author} {\bibfnamefont {H.~S.}\ \bibnamefont {Chu}},
  \bibinfo {author} {\bibfnamefont {V.~H.}\ \bibnamefont {Chen}}, \bibinfo
  {author} {\bibfnamefont {A.~Y.}\ \bibnamefont {Zhu}}, \ and\ \bibinfo
  {author} {\bibfnamefont {P.}~\bibnamefont {Genevet}},\ }\href@noop {}
  {\bibfield  {journal} {\bibinfo  {journal} {Optics Letters}\ }\textbf
  {\bibinfo {volume} {42}},\ \bibinfo {pages} {2639} (\bibinfo {year}
  {2017})}\BibitemShut {NoStop}%
\bibitem [{\citenamefont {Jafar-Zanjani}\ \emph {et~al.}(2018)\citenamefont
  {Jafar-Zanjani}, \citenamefont {Inampudi},\ and\ \citenamefont
  {Mosallaei}}]{jafar2018adaptive}%
  \BibitemOpen
  \bibfield  {author} {\bibinfo {author} {\bibfnamefont {S.}~\bibnamefont
  {Jafar-Zanjani}}, \bibinfo {author} {\bibfnamefont {S.}~\bibnamefont
  {Inampudi}}, \ and\ \bibinfo {author} {\bibfnamefont {H.}~\bibnamefont
  {Mosallaei}},\ }\href@noop {} {\bibfield  {journal} {\bibinfo  {journal}
  {Scientific Reports}\ }\textbf {\bibinfo {volume} {8}},\ \bibinfo {pages}
  {11040} (\bibinfo {year} {2018})}\BibitemShut {NoStop}%
\bibitem [{\citenamefont {Schmitt}\ \emph {et~al.}(2019)\citenamefont
  {Schmitt}, \citenamefont {Georg}, \citenamefont {Bri{\`e}re}, \citenamefont
  {Loukrezis}, \citenamefont {H{\'e}ron}, \citenamefont {Lanteri},
  \citenamefont {Klitis}, \citenamefont {Sorel}, \citenamefont {R{\"o}mer},
  \citenamefont {De~Gersem} \emph {et~al.}}]{schmitt2019optimization}%
  \BibitemOpen
  \bibfield  {author} {\bibinfo {author} {\bibfnamefont {N.}~\bibnamefont
  {Schmitt}}, \bibinfo {author} {\bibfnamefont {N.}~\bibnamefont {Georg}},
  \bibinfo {author} {\bibfnamefont {G.}~\bibnamefont {Bri{\`e}re}}, \bibinfo
  {author} {\bibfnamefont {D.}~\bibnamefont {Loukrezis}}, \bibinfo {author}
  {\bibfnamefont {S.}~\bibnamefont {H{\'e}ron}}, \bibinfo {author}
  {\bibfnamefont {S.}~\bibnamefont {Lanteri}}, \bibinfo {author} {\bibfnamefont
  {C.}~\bibnamefont {Klitis}}, \bibinfo {author} {\bibfnamefont
  {M.}~\bibnamefont {Sorel}}, \bibinfo {author} {\bibfnamefont
  {U.}~\bibnamefont {R{\"o}mer}}, \bibinfo {author} {\bibfnamefont
  {H.}~\bibnamefont {De~Gersem}},  \emph {et~al.},\ }\href@noop {} {\bibfield
  {journal} {\bibinfo  {journal} {Optical Materials Express}\ }\textbf
  {\bibinfo {volume} {9}},\ \bibinfo {pages} {892} (\bibinfo {year}
  {2019})}\BibitemShut {NoStop}%
\bibitem [{\citenamefont {Yang}\ and\ \citenamefont
  {Fan}(2017)}]{yang2017topology}%
  \BibitemOpen
  \bibfield  {author} {\bibinfo {author} {\bibfnamefont {J.}~\bibnamefont
  {Yang}}\ and\ \bibinfo {author} {\bibfnamefont {J.~A.}\ \bibnamefont {Fan}},\
  }\href@noop {} {\bibfield  {journal} {\bibinfo  {journal} {Optics Letters}\
  }\textbf {\bibinfo {volume} {42}},\ \bibinfo {pages} {3161} (\bibinfo {year}
  {2017})}\BibitemShut {NoStop}%
\bibitem [{\citenamefont {Piggott}\ \emph {et~al.}(2014)\citenamefont
  {Piggott}, \citenamefont {Lu}, \citenamefont {Babinec}, \citenamefont
  {Lagoudakis}, \citenamefont {Petykiewicz},\ and\ \citenamefont
  {Vu{\v{c}}kovi{\'c}}}]{piggott2014inverse}%
  \BibitemOpen
  \bibfield  {author} {\bibinfo {author} {\bibfnamefont {A.~Y.}\ \bibnamefont
  {Piggott}}, \bibinfo {author} {\bibfnamefont {J.}~\bibnamefont {Lu}},
  \bibinfo {author} {\bibfnamefont {T.~M.}\ \bibnamefont {Babinec}}, \bibinfo
  {author} {\bibfnamefont {K.~G.}\ \bibnamefont {Lagoudakis}}, \bibinfo
  {author} {\bibfnamefont {J.}~\bibnamefont {Petykiewicz}}, \ and\ \bibinfo
  {author} {\bibfnamefont {J.}~\bibnamefont {Vu{\v{c}}kovi{\'c}}},\ }\href@noop
  {} {\bibfield  {journal} {\bibinfo  {journal} {Scientific Reports}\ }\textbf
  {\bibinfo {volume} {4}},\ \bibinfo {pages} {7210} (\bibinfo {year}
  {2014})}\BibitemShut {NoStop}%
\bibitem [{\citenamefont {Callewaert}\ \emph {et~al.}(2018)\citenamefont
  {Callewaert}, \citenamefont {Velev}, \citenamefont {Kumar}, \citenamefont
  {Sahakian},\ and\ \citenamefont {Aydin}}]{callewaert2018inverse}%
  \BibitemOpen
  \bibfield  {author} {\bibinfo {author} {\bibfnamefont {F.}~\bibnamefont
  {Callewaert}}, \bibinfo {author} {\bibfnamefont {V.}~\bibnamefont {Velev}},
  \bibinfo {author} {\bibfnamefont {P.}~\bibnamefont {Kumar}}, \bibinfo
  {author} {\bibfnamefont {A.}~\bibnamefont {Sahakian}}, \ and\ \bibinfo
  {author} {\bibfnamefont {K.}~\bibnamefont {Aydin}},\ }\href@noop {}
  {\bibfield  {journal} {\bibinfo  {journal} {Scientific Reports}\ }\textbf
  {\bibinfo {volume} {8}},\ \bibinfo {pages} {1358} (\bibinfo {year}
  {2018})}\BibitemShut {NoStop}%
\end{thebibliography}%

%\bibliographystyle{apsrev4-1} 
%\bibliography{biblio} % mon fichier de base de donnees s'appelle
                      % biblio.bib

\end{document}